\documentclass[pra,onecolumn,floatfix,a4paper,superscriptaddress]{revtex4}
\usepackage{bm,color,graphicx,amsmath,txfonts}
\usepackage{here}
\usepackage[colorlinks, citecolor=blue,linkcolor=blue]{hyperref}
\usepackage{braket}

\newcommand{\e}{{e}}

\begin{document}

\makeatletter
\renewcommand{\@biblabel}[1]{\makebox[2em][l]{\textsuperscript{\textcolor{black}{\fontsize{10}{12}\selectfont[#1]}}}}
\makeatother

\let\oldbibliography\thebibliography
\renewcommand{\thebibliography}[1]{%
  \addcontentsline{toc}{section}{\refname}%
  \oldbibliography{#1}%
  \setlength\itemsep{0pt}%
}

\title{Exploring the hierarchy of quantum correlations under thermal effects in two gravitational cat states}

\author{Elhabib Jaloum}
\affiliation{LPTHE-Department of Physics, Faculty of sciences, Ibnou Zohr University, Agadir, Morocco}

\author{Mohamed Amazioug}
\affiliation{LPTHE-Department of Physics, Faculty of sciences, Ibnou Zohr University, Agadir, Morocco}

\begin{abstract}

In this article, we investigate the hierarchy of quantum correlations between two gravitational cats states (modeled by two qubits). We use concurrence to quantify the entanglement between the two gravitational cat states. Quantum steering is employed to measure the steerabilities. We consider geometric quantum discord to quantify quantum correlations beyond entanglement. We show that the concurrence persists even when steerability is lost under thermal effects. We also show that the temperature influences the degree of quantum correlations between the two gravitational cat states. Besides, when the energy difference between the ground state and the first excited level becomes significant, the states become separable.

\end{abstract}

\date{\today}

\maketitle

\section{Introduction}    \label{sec:1}

The quantum superposition \citep{1,2} plays a central role in how quantum information is represented and manipulated in quantum systems. It allows a qubit to simultaneously occupy multiple states. Moreover, quantum systems are characterized by certain correlations, such as quantum entanglement \citep{3,4} that cannot be defined using any classical means. Other phenomena include Bell nonlocality \citep{5}, quantum steering \citep{8,9}, and quantum discord \citep{6,7}. Quantum entanglement \citep{10} is a fundamental phenomenon in quantum physics and quantum technology \citep{11}. It refers to a state in which two particles can be entangled in such a way that the quantum state of each particle cannot be described independently of the other, even if they are separated by large distances. This means that a change in the state of one particle will instantly affect the other. While the transmission of information between two entangled qubits becomes possible, this capability can be exploited to build efficient quantum computers \citep{12,13} and develop associated technologies \citep{14,15}. Quantum steering is a phenomenon in quantum mechanics \citep{16} that involves the correlations between entangled particles \citep{17,18}, specifically in the context of measurements made on one particle influencing the state of another particle, even when the particles are spatially separated. First introduced by Erwin Schr\"odinger in response to the famous Einstein-Podolsky-Rosen (EPR) paradox \cite{19,20}. It is a form of quantum entanglement that exhibits a unique asymmetry: one party, referred to as the steerer (typically denoted as Alice), has the ability to influence or steer the state of another distant particle, often called the target (usually denoted as Bob), through the choice of measurements on her own entangled particle. Unlike Bell non-locality \cite{21} and quantum entanglement, in a steerable system, Alice has the ability to influence Bob's particle, but the reverse is not necessarily true. This unique characteristic of quantum steering has sparked significant interest in both theoretical and experimental quantum physics.\\

The Geometry of Quantum Discord (GQD) is a measure of quantum correlations in bipartite systems, designed to capture more general non-classical correlations beyond quantum entanglement. Geometric methods are frequently employed to define and measure quantum resources in a multitude of quantum systems \citep{72}. Specifically, the Schatten 1-norm quantum discord \citep{73,74}, is a reliable geometric-based indicator used to assess the level of quantum correlations in metal complexes \citep{75}. Unlike other correlation measures, it takes into account the geometry of the quantum state space. This approach allows quantifying quantum correlations in a way that considers the geometric structure of the quantum state space, which can provide richer information about quantum interactions. GQD has been widely studied both theoretically and experimentally, offering interesting prospects for the characterization and utilization of quantum resources in various contexts of quantum information processing and quantum communication \citep{44,45}.
On the other hand, quantum coherence is a fundamental concept in quantum physics. It refers to the ability of a quantum system to maintain a superposition state, where multiple possibilities of states exist simultaneously. Moreover, quantum coherence is crucial for understanding quantum phenomena such as interference and other aspects of quantum optics \citep{46,47}. In various research domains, quantum coherence has become a valuable resource. Among these, it is used to study quantum thermodynamics \citep{48,49}, quantum dots \citep{50,51}, Heisenberg spin chains \citep{52,53}, spin waves \citep{54}, quantum batteries \citep{55}, and even to develop algorithms for quantum computing \citep{56,57}.These applications highlight the importance of quantum coherence in many areas of quantum physics and its technological applications. In other words, when attempting to observe such phenomena on a macroscopic scale, quantum coherence tends to rapidly degrade due to interaction with the environment, a phenomenon known as decoherence. Quantum decoherence \cite{29} is a process by which a quantum system loses its coherence and quantum character, behaving more like a classical system. When entangled particles interact with their environment, interactions with surrounding particles result in a loss of quantum coherence \citep{30}, which can disrupt or destroy entanglement between these particles. However, in some cases, decoherence can also play a role in protecting entanglement, depending on the specific details of the interaction between the quantum system and its environment.\\

At the subatomic scale, a stationary heavy particle can be in a Schr\"dinger cat state or in a superposition of two physically distinct positions. These states are called gravitational entities and are known as gravitational cats (gravcats) \cite{22}. The quantification of gravitational interaction refers to the effort to develop a quantum description of gravity. Albert Einstein's General Theory of Relativity provides a classical description of gravity, explaining how mass and energy curve space-time, thus influencing the trajectories of moving objects. However, integrating it with quantum physics, which describes phenomena at the microscopic scale, poses significant conceptual challenges. Some approaches include string theory \cite{23}, loop quantum gravity \cite{24,25}, and other quantum gravity theories. The development of a theory of quantum gravity is crucial for our ultimate understanding of the universe, especially in extreme situations such as black holes \citep{26,27} and the Big Bang \citep{28}, where quantum and gravitational effects overlap.\\

In this paper, we investigate quantum correlations (quantum steering, quantum concurrence, and geometric quantum discord) between two gravitational cat states (gravcats). We use concurrence to quantify the entanglement, quantum steering to quantify the steerabilities, and GQD to measure quantum correlations beyond the entanglement between the two qubits. We prove that quantum entanglement remains more persistent than quantum steering in the system under consideration. We show that quantum correlations depend on the temperature, potential energy and  excitation energy.\\

This article is structured as follows: in Section \ref{sec:2} we present the model describing the gravitational interaction that generates two qubits (which we call gravcat states), and we introduce the thermal density operator due to the bath thermal. In Section \ref{sec:3}, we present quantum steering, quantum entanglement, and geometric quantum discord. In Section \ref{sec:4}, we present quantum steering as a function of temperature and different model parameters, and we compare different quantum correlations such as quantum steering, concurrence, and geometric quantum discord. Finally, in Section \ref{sec:5}, we present our conclusions and further discussions.

\section{Hamiltonian and Gibb's density operator}
\label{sec:2}

At the Planck scale, a particle of mass $m$ can exist in a superposition of two states (Schrodinger's cat state). The model under study consists of two particles, each of which is contained within a one-dimensional double-well potential. Let's consider a particle of mass $m$, each corresponding to a qubit with minima at $\hat{x}=\pm \frac{L}{2}$, with the Hamiltonian $\hat{H}=\frac{1}{2m}\hat{p}^{2}+U(\hat{x})$,we denote the ground state of the particle as $\ket{f}$ and $\ket{e}$ as the first excited state, and the energy difference between these states is $\omega$
\begin{equation}
\ket{\pm}=\frac{1}{\sqrt{2}}(\ket{f}\pm\ket{e})
\end{equation}
We define the states $\ket{0}$ and $\ket{1}$ according to the Landau-Lifschitz approximation as \citep{31}
\begin{equation}
\ket{\pm}=\frac{1}{\sqrt{2}}\left( \ket{0}\pm\ket{1}\right)
\end{equation}
In the non-relativistic limit, the Hamiltonian describing the gravitational interaction between two gravcats \cite{31}
\begin{equation}
\hat{H}=\frac{\omega}{2}(\sigma_{z}\otimes\mathbb{I}+\mathbb{I}\otimes \sigma_{z})-\gamma(\sigma_{x}\otimes\sigma_{x})
\end{equation}
where $\sigma_{x,z}$ represent the familiar Pauli matrices, acting as operators on the internal state of each gravcat and $\omega$ captures the energy difference between the ground and excited states of each individual gravcat. We introduce a parameter, $\gamma$, that measures the intensity of the gravitational interaction between the two gravcats, where $\gamma=\frac{ G m^{2}}{2}(\frac{1}{d_{1}}-\frac{1}{d_{2}})$ with $G$ is a universal gravitational constant and the terms $d_{1}$ and $d_{2}$ represent the equilibrium separations between the two gravcats. These are not just any distances, but rather the specific distances where the gravitational force on each gravcat is minimized.
\begin{figure}[H]
\begin{center}
\includegraphics[width=9cm,height=10cm]{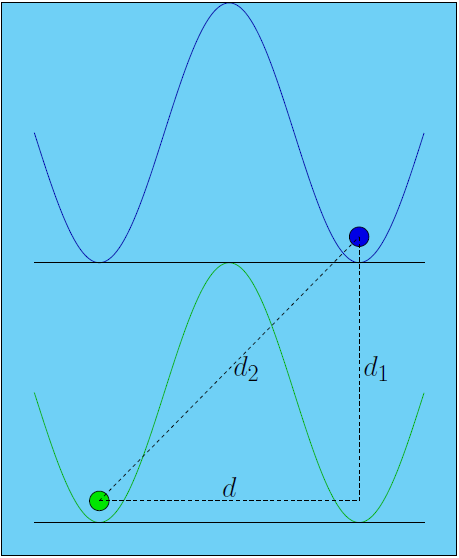}
\caption{Schematic of two interacting gravcats situated within an even double-well potential. The potential has two minimal energy regions, separated by a distance $d$, where $d_{2}=\sqrt{d_{1}^{2}+d^{2}}$.}
\end{center}
\end{figure}
The calculation of the preceding Hamiltonian yields
\begin{equation}
\hat{H}=
\begin{pmatrix}
\omega & 0 & 0 & -\gamma \\
0 & 0 & -\gamma & 0 \\
0 & -\gamma & 0 & 0 \\
-\gamma & 0 & 0 & -\omega
\end{pmatrix}
\end{equation}
Once we diagonalize the Hamiltonian, we arrive at a set of specific values
\begin{equation}
    \left\{
    \begin{array}{ll}
\lambda_{1,2}=\mp\gamma\\
\lambda_{3,4}=\mp\sqrt{\gamma^{2}+\omega^{2}}
 \end{array}
    \right.
\end{equation}
    
and associated eigenvectors
\begin{equation}
    \left\{
    \begin{array}{ll}
\vspace*{0.2cm} 
\ket{\psi_{1}}=\frac{1}{\sqrt{2}}\left( \ket{01}+\ket{10}\right)  \\
\vspace*{0.2cm} 
\ket{\psi_{2}}=\frac{1}{\sqrt{2}}\left( \ket{01}-\ket{10} \right)\\
\vspace*{0.2cm} 
\ket{\psi_{3}}=\cos(\kappa_{+})\ket{00}+sin(\kappa_{+})\ket{11}\\
\vspace*{0.2cm} 
\ket{\psi_{4}}=\cos(\kappa_{-})\ket{00}+sin(\kappa_{-})\ket{11}\\
\end{array}
    \right.
\end{equation}    
where $\kappa_{\pm}$ are defined by
\[
\kappa_{\pm}=\arctan\left(  \frac{\gamma}{\omega\pm\sqrt{\omega^{2}+\gamma^{2}}}\right) 
\]
In thermal equilibrium with a heat bath, the state of the system being in a particular state is writes as
\begin{equation}
\varrho(T)=\frac{\e^{-\beta H}}{Z}
\end{equation}
where  $Z=Tr[\exp({-\beta H})]$ and $\beta=\frac{1}{k_{B}T}$,where $k_{B}$ is the Boltzmann constant and $T$ is the absolute temperature. This system's thermal state is fully characterized by the quantum density operator $\varrho$ as
\begin{equation}
\varrho=
\begin{pmatrix}
\varrho_{1,1}& 0 & 0 & \varrho_{1,4} \\
0 & \varrho_{2,2} & \varrho_{2,3} & 0 \\
0 & \varrho_{3,2} & \varrho_{3,3}& 0 \\
\varrho_{4,1} & 0 & 0 & \varrho_{4,4}
\end{pmatrix},
\label{eq:T}
\end{equation}
where
$$
\begin{aligned}
\phantom{h(x) }\varrho_{1,1}&=\frac{\cos^{2}(\kappa_{+})\exp(-\beta\lambda_{3})+\cos^{2}(\kappa_{-})\exp(-\beta\lambda_{4})}{Z},\\
\phantom{h(x) }\varrho_{2,2}&=\varrho_{3,3}=\frac{\exp(-\beta\lambda_{1})+\exp(-\beta\lambda_{2})}{2Z},\\
\phantom{h(x) }\varrho_{2,3}&=\varrho_{3,2}=\frac{\exp(-\beta\lambda_{1})-\exp(-\beta\lambda_{2})}{2Z},\\
\phantom{h(x) }\varrho_{1,4}&=\varrho_{4,1}=\frac{\sin(2\kappa_{+})\exp(-\beta\lambda_{3})+\sin(2\kappa_{-})\exp(-\beta\lambda_{4})}{2Z},\\
\phantom{h(x) }\varrho_{4,4}&=\frac{\sin^{2}(\kappa_{+})\exp(-\beta\lambda_{3})+\sin^{2}(\kappa_{-})\exp(-\beta\lambda_{4})}{Z},
\end{aligned}
$$

Where $\lambda_{i}$ are the eigenenergies of the Hamiltonian $H$, corresponding to the eigenstates $\psi_{i}$, and 
\[Z=2\cosh\left[ \beta\gamma\right] +2\cosh\left[ \sqrt{\beta(\omega^{2}+\gamma^{2})}\right] \]
assumes this form for the diagonalized
Hamiltonian. The X-shaped form of the thermal density operator arises from the symmetry of the Hamiltonian.   Additionally, for validity, the thermal density operator must be Hermitian. Furthermore, $\varrho$ must satisfy that $Tr(\varrho)=1$ and $Tr(\varrho^{2})\leq 1$,  given that we are dealing with a mixed density operator. Since $\varrho$ describes a thermal state, the correlations of this system are referred to as thermal quantum correlations.
\section{Quantum correlation measures}
\label{sec:3}
The indirect detection of Bell non-locality through quantum steering \cite{32}, and the detection of quantum steering by quantum entanglement \cite{33}, constitutes an intriguing concept in quantum physics. Quantum steering is a phenomenon in which the measurement performed on a particle at one location $A$ seems to instantaneously influence the state of another particle at a distant location $B$.
\subsection{Quantum steering}
Let us imagine that Alice and Bob share quantum bonds characterized by the density matrix of their two qubits, $\varrho_{AB}$. Alice's objective is to demonstrate to Bob that she can influence the state of his qubit by performing measurements on his. For a general bipartite state $\varrho_{AB}$ shared by Alice and Bob, the phenomenon of steering from Bob to Alice (or Alice to Bob) can be observed if the density matrix $\varrho_{B\rightarrow A}$ (or $\varrho_{A\rightarrow  B})$ is defined as follows \cite{32,34}
\begin{equation}
\varrho_{A\rightarrow B}=\frac{1}{\sqrt{3}}\varrho_{AB}+\left(1-\frac{1}{\sqrt{3}}\right)\hat{\varrho}_{B}
\label{eq:s}
\end{equation}
where $\hat{\varrho}_{B}=\frac{\mathbb{I}}{2}\otimes \varrho_{B}$ with $\mathbb{I}$ two-dimensional identity matrix and $\varrho_{B}=Tr_{A}(\varrho_{AB})$ being the reduced state at Bob's side.\\
Furthermore, for any qubit-qubit state $\varrho_{AB}$ shared between Alice and Bob, let's define another new qubit-qubit state $\varrho_{B\rightarrow A}$ as \citep{35,36}
\begin{equation}
\varrho_{B\rightarrow A}=\frac{1}{\sqrt{3}}\varrho_{AB}+\left(1-\frac{1}{\sqrt{3}}\right)\hat{\varrho}_{A}
\end{equation}
where $\hat{\varrho}_{A}=\varrho_{A}\otimes\frac{\mathbb{I}}{2}$ with $\mathbb{I}$ two-dimensional identity matrix, and $\varrho_{A}=Tr_{B}(\varrho_{AB})$ being the reduced state at Alice’s side.\\
Through simple calculations,the matrix $\varrho_{B\rightarrow A}$ of the X-state $\rho_{AB}$ can be expressed as
\begin{equation}
\varrho_{B\rightarrow A}=
\begin{pmatrix}
\frac{\sqrt{3}}{3}\varrho_{1,1}+a& 0 & 0 & \frac{\sqrt{3}}{3}\varrho_{1,4} \\
0 &\frac{\sqrt{3}}{3}\varrho_{2,2}+a& \frac{\sqrt{3}}{3}\varrho_{2,3}& 0 \\
0 & \frac{\sqrt{3}}{3}\varrho_{3,2} & \frac{\sqrt{3}}{3}\varrho_{3,3}+b& 0 \\
\frac{\sqrt{3}}{3}\varrho_{4,1} & 0 & 0 & \frac{\sqrt{3}}{3}\varrho_{4,4}+b
\end{pmatrix}
\end{equation}
where $a=\frac{3-\sqrt{3}}{6}(\varrho_{1,1}+\varrho_{2,2})$ and $b=\frac{3-\sqrt{3}}{6}(\varrho_{3,3}+\varrho_{4,4})$\\
The density matrices $\varrho_{AB}$ constructed thus preserve the X structure. Using Equation (\ref{eq:s}), the state $\varrho_{A\rightarrow B}$  is entangled if the state $\varrho_{A\rightarrow B}$ satisfies the inequality \cite{4}
\begin{subequations}
\begin{align}
|\varrho_{1,4}|^{2}>f_{a}-f_{b}
\end{align}
\begin{align}
|\varrho_{2,3}|^{2}>f_{c}-f_{b}
\end{align}
\end{subequations}
where:
\begin{subequations}
\begin{align}
f_{a}=\frac{2-\sqrt{3}}{2}\varrho_{1,1}\varrho_{4,4}+\frac{2+\sqrt{3}}{2}\varrho_{2,2}\varrho_{3,3}+\frac{1}{4}(\varrho_{1,1}+\varrho_{4,4})(\varrho_{2,2}+\varrho_{3,3}).
\end{align}
\begin{align}
f_{b}=\frac{1}{4}(\varrho_{1,1}-\varrho_{4,4})(\varrho_{2,2}-\varrho_{3,3}).
\label{eq:b}
\end{align}
\begin{align}
f_{c}=\frac{2+\sqrt{3}}{2}\varrho_{1,1}\varrho_{4,4}+\frac{2-\sqrt{3}}{2}\varrho_{2,2}\varrho_{3,3}+\frac{1}{4}(\varrho_{1,1}+\varrho_{4,4})(\varrho_{2,2}+\varrho_{3,3}).
\end{align}
\end{subequations}
Using a similar method, we find that the direction from Alice to Bob can be verified by one of the following inequalities:
\begin{center}
\begin{subequations}
\begin{align}
|\varrho_{1,4}|^{2}>f_{a}+f_{b}
\end{align}
\begin{align}
|\varrho_{2,3}|^{2}>f_{c}+f_{b}
\end{align}
\end{subequations}
\end{center}
According to the inequality, the Steerability of Bob to Alice $S_{B\rightarrow A}$ is equal to  \cite{32}
\begin{equation}
S_{B\rightarrow A}=\max\left\lbrace 0,\frac{8}{\sqrt{3}}[|\varrho_{1,4}|^{2}-f_{a}+f_{b},|\varrho_{2,3}|^{2}-f_{a}+f_{b}]\right\rbrace
\end{equation}
By exchanging mode $A$ and mode $B$, we can obtain the steerability $S_{A\rightarrow B}$ as follows
\begin{equation}
S_{A\rightarrow B}=\max\left\lbrace 0,\frac{8}{\sqrt{3}}[|\varrho_{1,4}|^{2}-f_{a}-f_{b},|\varrho_{2,3}|^{2}-f_{a}-f_{b}]\right\rbrace 
\end{equation}

The steering asymmetry is defined as the absolute difference between the steering measures from Alice to Bob $S_{A\rightarrow B}$ and from Bob to Alice $S_{B\rightarrow A}$
\begin{equation}
\Delta_{12}=\vert S_{A\rightarrow B}-S_{B\rightarrow A} \vert
\end{equation}
The steering relationships between qubit 1 held by Alice $A$ and qubit 2 held by Bob $B$ can vary in several ways.\\
When $\Delta_{12}>0$ (one-way steering), it indicates an asymmetry in the ability of both parties to influence each other's quantum states. Specifically, if $S_{A\rightarrow B}>0$ and $S_{B\rightarrow A}=0$, this reveals that Alice can influence Bob's qubit while Bob cannot influence Alice's qubit, thus defining one-way steering from Alice to Bob. Similarly, if $S_{A\rightarrow B}=0$ and $S_{B\rightarrow A}>0$, it means that Bob can influence Alice's qubit without Alice being able to influence Bob's qubit, also presenting a form of one-way steering, but this time from Bob to Alice. In the case where $\Delta_{12}=0$ (two-way steering or no steering), this indicates that there is no asymmetry in the ability of the parties to mutually influence quantum states. Thus, both parties, Alice and Bob, can affect each other's quantum states equally, characterizing two-way steering. Additionally, this situation may also imply that there is no steering at all, suggesting that both parties cannot significantly influence each other's quantum states.

\subsection{Quantum entanglement}
As we all know, quantum entanglement of the bipartite states can be effectively identified by the concurrence. The quantum entanglement of bipartite states can be effectively identified through concurrence. Let's use Wootters concurrence to determine the concurrence of the state $\varrho_{AB}$ \citep{37,42}
\begin{center}
\begin{equation}
C(\varrho_{AB})=\max\left\lbrace  \sqrt{\xi_{1}}-\sqrt{\xi_{2}}-\sqrt{\xi_{3}}-\sqrt{\xi_{4}},0\right\rbrace  
\label{eq:C}
\end{equation}
\end{center}
The $\xi_{i}$ represent the eigenvalues in decreasing order of the matrix $\hat{\varrho}_{AB}$.Note that each $\xi_{i}$ is a non-negative real number
\begin{center}
\begin{equation}
\hat{\varrho}=\varrho_{AB}(\sigma_{y}\otimes\sigma_{y})\varrho_{AB}^{*}(\sigma_{y}\otimes\sigma_{y})
\end{equation}
\end{center}
where $\varrho^{*}$ denotes the complex conjugation of $\varrho$ and $\sigma_{y}$ is a Pauli matrix.\\
The concurrence of the state can be obtained from Equation (\ref{eq:C}), resulting in
\begin{center}
\begin{equation}
C(\varrho_{AB})=2\max \left\lbrace |\varrho_{2,3}|-\sqrt{\varrho_{1,1}\varrho_{4,4}},|\varrho_{1,4}|-\sqrt{\varrho_{2,2}\varrho_{3,3}},0\right\rbrace .
\end{equation}
It is essential to underline that the concurrence $0\leq C(\varrho_{AB})\leq 1$,the minimum and maximum values indicate, respectively, separable states and maximally entangled states.
\end{center}
\subsection{Geometric Quantum Discord}
The quantum discord provides a more comprehensive measure of quantum correlations in a system, taking into account all forms of quantum correlation. It was initially introduced as an entropy-based measure that assesses the actual quantum correlations present in a quantum state \citep{62}. In this measure, the mutual information between subsystems A and B is represented by 
\begin{equation}
Q(\rho_{AB})=I(\rho_{A}:\rho_{B})-C(\rho_{AB})
\end{equation}
where, $I(\rho_{A}:\rho_{B})=S(\rho_{A})+S(\rho_{B})-S(\rho_{AB})$ represents the mutual information between subsystems A and B, while $C(\rho_{AB})$ denotes the classical correlation of the composite system $\rho_{AB}$, defined as $C(\rho_{AB})=\max_{B_{k}}[S(\rho_{A}-\sum_{k} p_{k}S(\rho_{k})]$, This involves maximizing over positive operator-valued measurements POVM $B_{k}$ applied exclusively to subsystem $B$. However, even for a two-qubit system \citep{63,64}, performing this analytical maximization over POVMs is a complex task. Faced with this challenge, various alternative approaches have been developed to assess quantum correlations \citep{65,66}. These methods incorporate geometric considerations rather than relying solely on entropic calculations \citep{67,68}, thus providing alternative options to characterize the quantum correlations present in quantum systems.\\
Geometric approaches are widely employed to assess and quantify quantum resources across various quantum systems, with particular emphasis on the Schatten 1-norm (or trace norm) quantum discord \citep{63,69}. The geometric quantum discord, for a two-qubit state ,can be expressed as \citep{70,71}
\begin{equation}
Q_{G}(\varrho_{AB})=\min_{\varrho_{c} \in \Omega} ||\varrho_{AB} - \varrho_{c}||_1
\label{eq:QG}
\end{equation}
where  $||\varrho_{AB} - \varrho_{c}||_1=Tr\sqrt{(\varrho_{AB} - \varrho_{c})^{\dagger}(\varrho_{AB} - \varrho_{c})}$ represents the 1-norm, also known as the trace norm, and $\varrho_{AB}$ denotes the quantum state at thermal equilibrium. The minimal distance between a set $\Omega$ of closest classical-quantum states $\varrho_{c}$ is given by
\begin{equation}
 \varrho_{c} = \sum_{k} p_{k} \Pi_{k,1} \otimes \rho_{k,2}
\end{equation}
where $0\leq p_{k}\leq 1 $ and $\sum_{k} p_{k}=1$, $\left\lbrace p_{k}\right\rbrace$ is a probability distribution; the orthogonal projectors associated with qubit 1 are denoted by $\Pi_{k,1}$, and the density matrix associated with the second qubit is $\rho_{k,2}$.\\
The minimization solution in Equation (\ref{eq:QG}) for a generic two-qubit X state enables us to express the trace quantum discord in the state $\varrho$, as indicated in Equation (\ref{eq:T}), thus providing the geometric quantum discord for the two-qubit X state, based on the Schatten 1-norm.

\begin{equation}
Q_{G}(\varrho_{AB})=\frac{1}{2}\sqrt{\frac{R_{1,1}^{2}\max\left\lbrace R_{2,2}^{2}+R_{3,0}^{2},R_{3,3}^{2}\right\rbrace -R_{2,2}^{2}\min\left\lbrace R_{1,1}^{2},R_{3,3}^{2}\right\rbrace }{\max\left\lbrace R_{2,2}^{2}+R_{3,0}^{2},R_{3,3}^{2}\right\rbrace-\min\left\lbrace R_{1,1}^{2},R_{3,3}^{2}\right\rbrace +R_{1,1}^{2}-R_{2,2}^{2}}}
\label{eq:Q}
\end{equation}
where $R_{\mu,\nu}$ in Equation (\ref{eq:Q}) being the components of the correlation matrix occurring after decomposing the state $\hat{\rho_{AB}}$ in the Fano-Bloch representation as
\begin{equation}
R=\sum_{\mu,\nu}R_{\mu,\nu}\sigma_{\mu}\otimes\sigma_{\nu}
\end{equation}
where the non-vanishing matrix elements $R_{\mu,\nu}$, are given by
$$
\begin{aligned}
\phantom{h(x) }R_{1,1} &=Tr\left[\sigma_{1}\otimes\sigma_{1}\varrho_{AB}\right] = 2\left(\varrho_{2,3}+\varrho_{1,4}\right) ,\\
\phantom{h(x) }R_{2,2} &=Tr\left[\sigma_{2}\otimes\sigma_{2}\varrho_{AB}\right]=2\left(\varrho_{2,3}-\varrho_{1,4}\right),\\
\phantom{h(x)}R_{3,3} &=Tr\left[\sigma_{3}\otimes\sigma_{3}\varrho_{AB}\right]=1-2\left( \varrho_{2,2}+\varrho_{3,3}\right),\\
\phantom{h(x)}R_{0,3} &=Tr\left[\sigma_{0}\otimes\sigma_{3}\varrho_{AB}\right]=2\left( \varrho_{1,1}-\varrho_{3,3}\right) -1,\\
R_{3,0} &=Tr\left[\sigma_{3}\otimes\sigma_{0}\varrho_{AB}\right]=2\left( \varrho_{1,1}-\varrho_{2,2}\right) -1,
\end{aligned}
$$

\section{Résultats and discussion}   \label{sec:4}
\begin{figure}[H]
\begin{center}
\includegraphics[width=18cm,height=8cm]{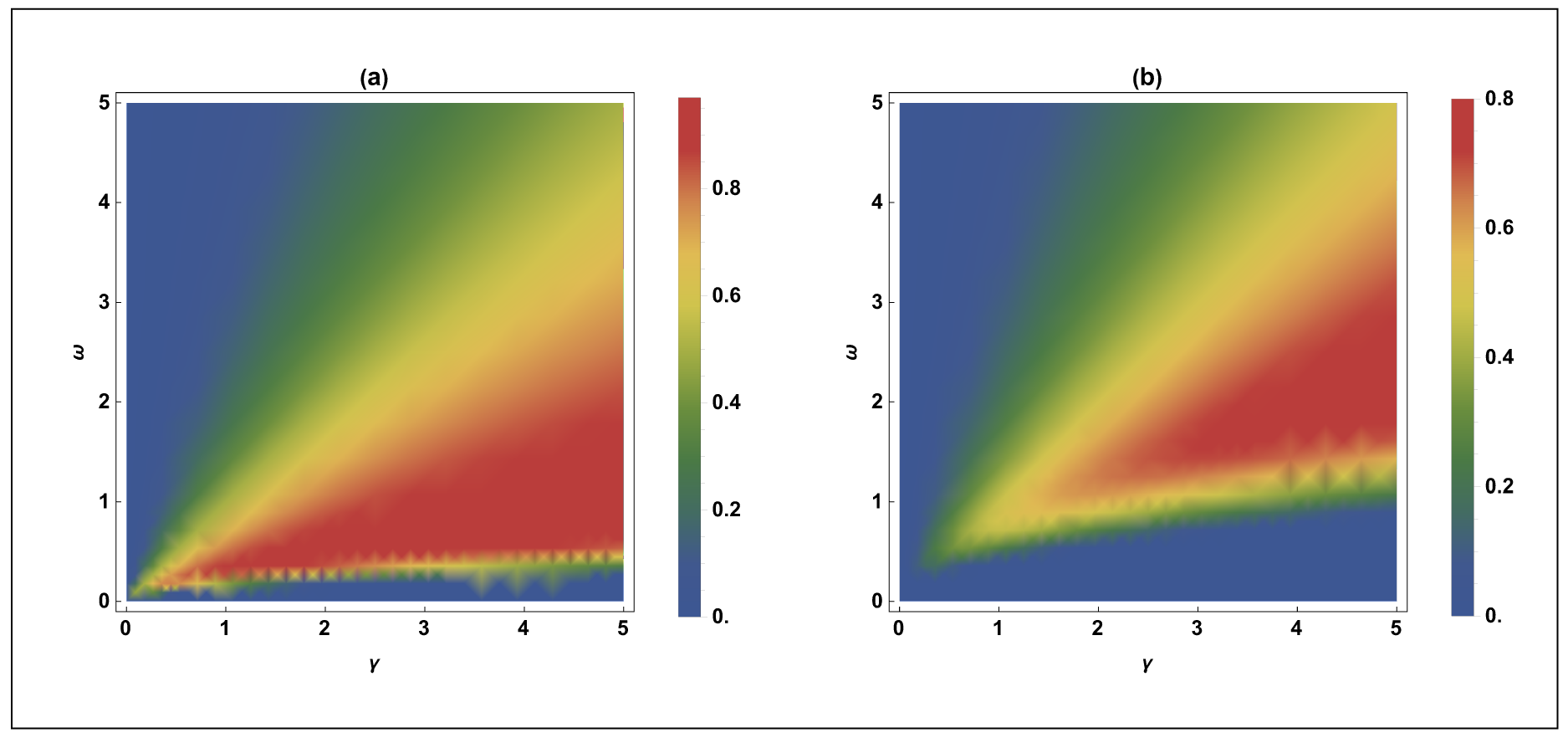}
\end{center}
\caption{Quantum steering in two gravcats versus $\omega$ and $\gamma$ with (a) $T=0.01$ and (b) $T=0.1$}
\label{fig:3}
\end{figure}

Here we delve into the two-state gravitational cat model using numerical methods to explore its properties. First, we will study the behavior of quantum steerability as a function of temperature $T$, potential energy $\gamma$ and excitation energy $\omega$. And as $\rho_{2,2}=\rho_{3,3}$, the inequality $f_{b}=0$ (Equation (\ref{eq:b})). Then the steerability of $A$ to $B$ becomes equal to that of $B$ to $A$ (two-way steering),
\begin{equation}
S_{A\rightarrow B}=S_{B\rightarrow A}=\max\left\lbrace 0,\frac{8}{\sqrt{3}}[|\rho_{1,4}|^{2}-f_{a},|\rho_{2,3}|^{2}-f_{c}]\right\rbrace
\end{equation}
and 
\begin{equation}
\Delta_{12}=0
\end{equation}
\begin{figure}[H]
\begin{minipage}{0.33\linewidth}
\includegraphics[scale=0.33]{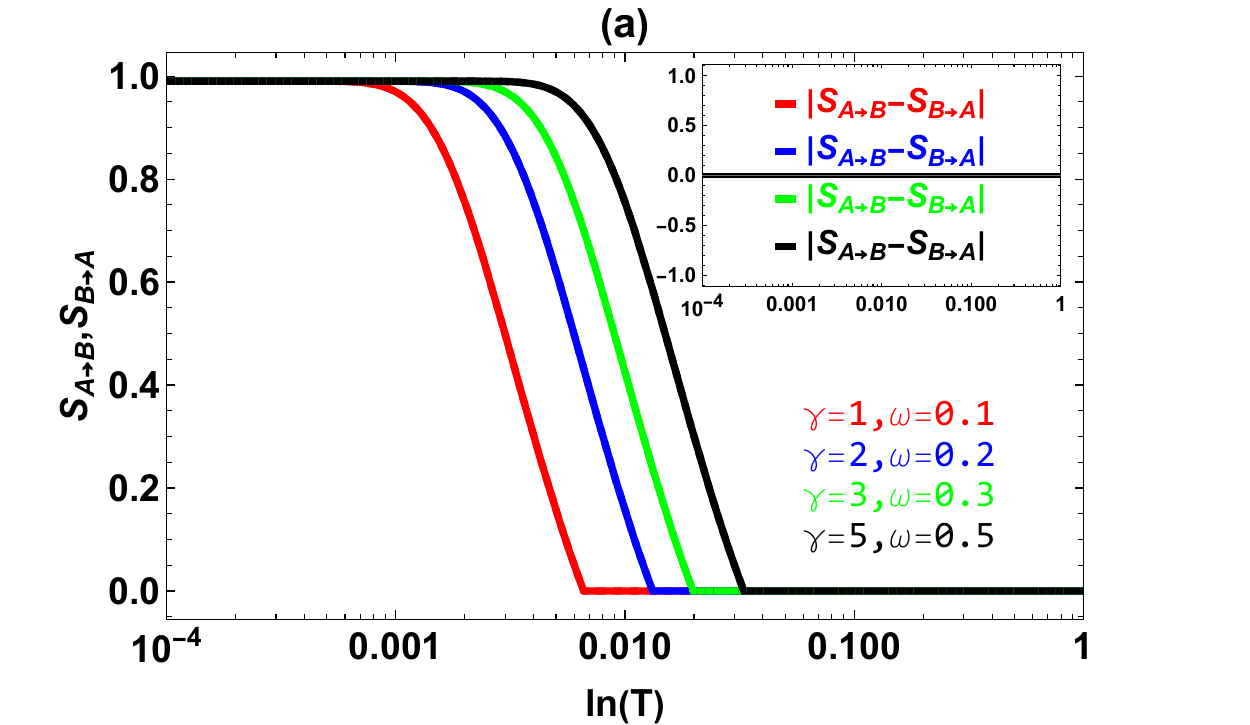}
\end{minipage}\hfill
\begin{minipage}{0.33\linewidth}
\includegraphics[scale=0.33]{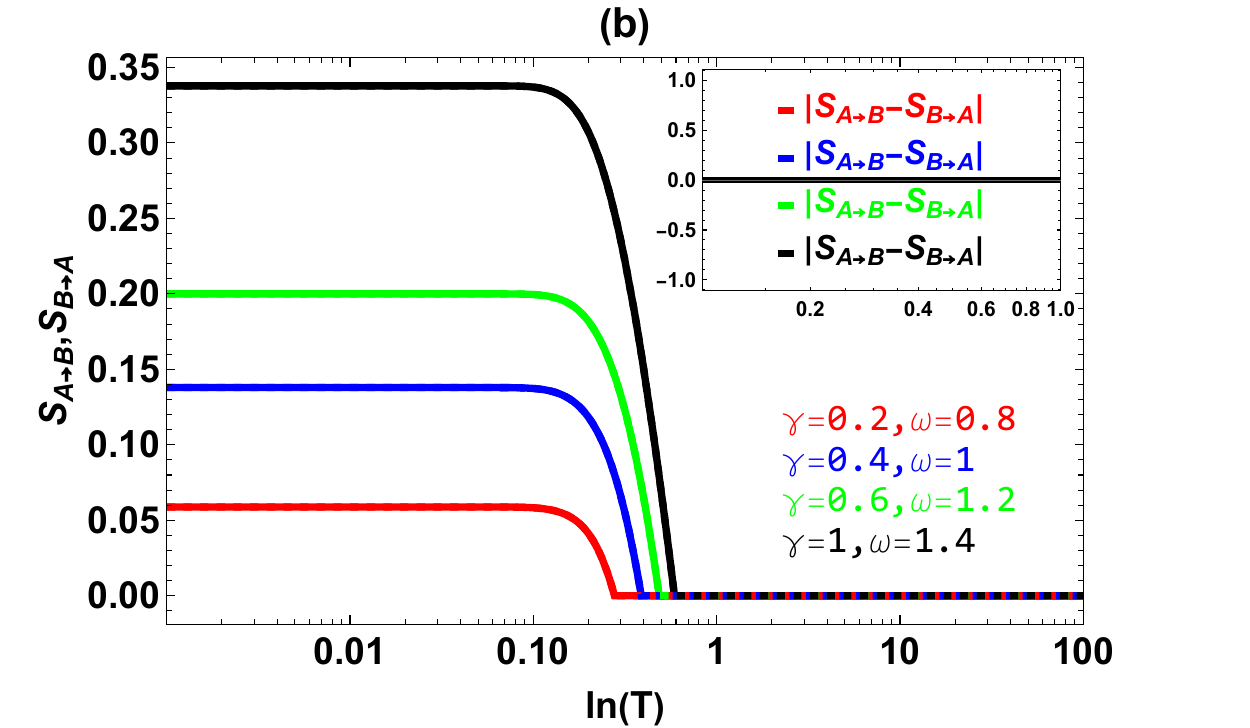}
\end{minipage}\hfill
\begin{minipage}{0.33\linewidth}
\includegraphics[scale=0.33]{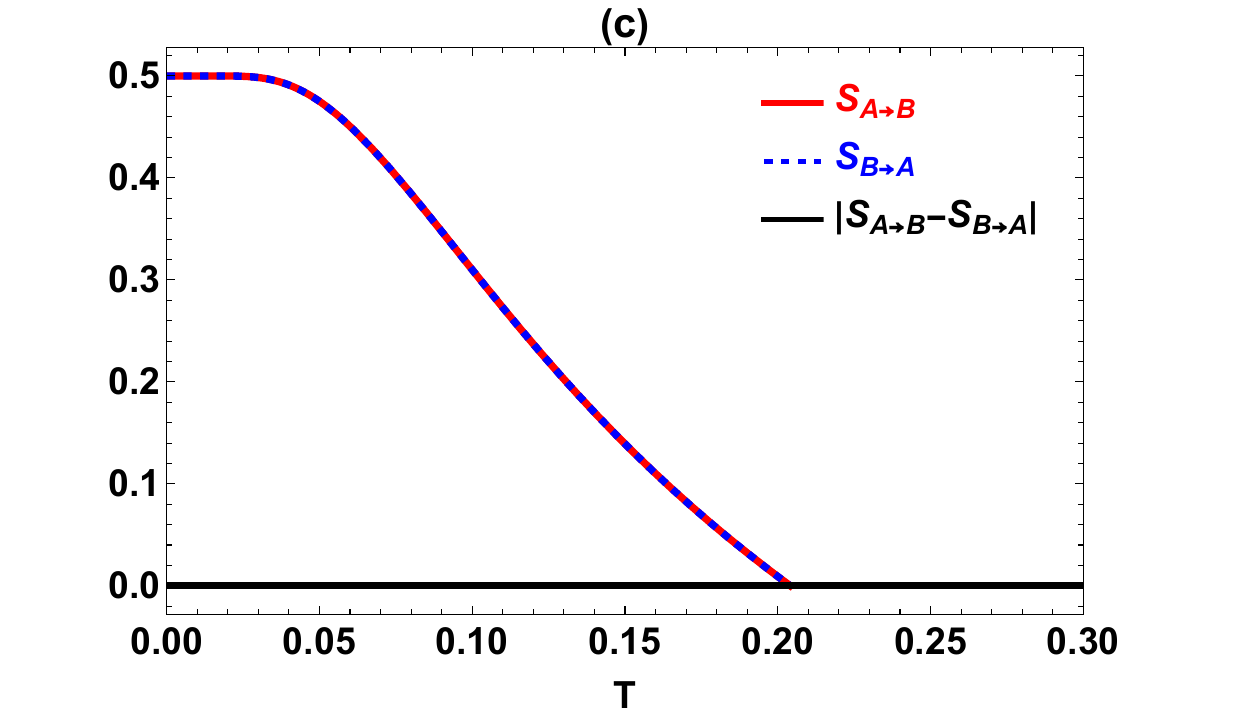}
\end{minipage}
\caption{(Color online) Plot of quantum steering $S_{A\rightarrow B}$ from qubit $A$ to qubit $B$ (solid curve) and $S_{B\rightarrow A}$ from qubit B to qubit $A$ (dashed curve) as a function of temperature in the logarithmic scale, for different values of $\gamma$ and $\omega$ such that (a): $\gamma>\omega$,(b):$\gamma<\omega$, (c): comparison of the quantum steering $S_{A\rightarrow B}$ and $S_{B\rightarrow A}$ for  $\gamma=\omega=0.5$}
\label{fig:2}
\end{figure}
In Fig.\ref{fig:2}(a), we plot the quantum steering versus the temperature $T$ with $\gamma =1$ and $\omega =0.1$ (red curve), $\gamma =2$ and $\omega =0.2$ (blue curve), $\gamma =3$ and $\omega =0.3$ (green curve), $\gamma = 5$ and $\omega =0.5$ (black curve). The parts of the system are symmetrical, and their ability to influence each other is equivalent, so $S_{A\rightarrow B}$ and $S_{B\rightarrow A}$ are equal ($S_{A\rightarrow B}=S_{B\rightarrow A}$). We notice that the steerability reaches its maximum value when the potential energy increases compared to the excitation energy. Moreover, steerability decreases when the temperature increases. This can be explained by the decoherence phenomenon. Besides, we remark that the threshold temperature, at which entanglement disappears, also rises with $\gamma$ and $\omega$. This remark helps elucidate the relation between the required temperature, mass, distance, and energy scales in experiments aiming to measure this effect.\\
In Fig.\ref{fig:2}(b), the quantum steering is represented as a function of the temperature $T$ in the logarithmic scale, at a fixed value of $\gamma =0.2$ and $\omega =0.8$ (red curve), $\gamma =0.4$ and $\omega =1$ (blue
curve), $\gamma =0.6$ and $\omega =1.2$ (green curve),$\gamma =1$ and $\omega =1.4$ (black curve). One can observe that, when $\gamma<\omega$, the quantum steering is weak: $S_{A\rightarrow B}=S_{B\rightarrow A}\approx 0.06$ (red curve), $S_{A\rightarrow B}=S_{B\rightarrow A}=0.14$ (blue curve), $S_{A\rightarrow B}=S_{B\rightarrow A}=0.2$ (green curve) and $S_{A\rightarrow B}=S_{B\rightarrow A}=0.34$ (black curve). We have $S_{A\rightarrow B}=S_{B\rightarrow A}$ (i.e. $\Delta_{12}=0$), this implies that the parts of the system are symmetrical. we notice that when $\gamma<\omega$, the steerability is low, in addition, we see that the quantum steering decreases with the increase in the temperature $T$. These results are consistent with expectations, as they demonstrate that the greater the mass of the state for fixed distances  or the energy scale of the model is elevated, the higher the intensity of gravity-mediated steerability. Similar behavior is observed for smaller distances between particles. However, interactions with the environment, such as decoherence and noise, can lead to undesirable effects that limit steerability, even in the presence of increased potential energy and excitement. It is important to consider other factors and specific conditions that could modulate this relationship in particular circumstances.\\
We plot in Fig.\ref{fig:2}(c) the steearbilities $S_{A\rightarrow B}$, $S_{B\rightarrow A}$, and the asymmetry $\Delta_{12}$ versus the temperature $T$. We note that $S_{A\rightarrow B}=S_{B\rightarrow A}>0$ (i.e., $\Delta_{12}=0$) when $T\leq  0.2$. This witnesses the existence of two-way steering between qubit 1 and qubit 2. Moreover, when $T\geq 0.2$, we have $S_{A\rightarrow B}=S_{B\rightarrow A}=0$. This means no-way steering between qubit 1 and qubit 2 (i.e., $\Delta_{12}=0$).\\
In Fig.\ref{fig:3}, we present the steerability as a function of $\gamma$ and $\omega$, comparing its properties for two temperatures: $T=0.01$ in Fig.\ref{fig:3}(a) and $T=0.1$ in Fig. \ref{fig:3}(b). For  the lower temperature ($T=0.01$), we notice that the steerability does not evolve linearly with $\gamma$ and $\omega$. For instance, if we fix a specific value of $\gamma$, we may observe an increase in steerability with $\omega$. However, this trend may change for other values of $\gamma$ and $\omega$, rendering the overall behavior non-monotonic. On the other hand, for a higher temperature ($T=0.1$),  the steerability initially decreases for relatively low values of $\gamma$ and $\omega$. This is due to thermal effects introducing more noise and reducing coherence in the system. However, as $\gamma$ increases, the steerability tends to improve due to stronger correlations among the particles in the system. Similarly, increasing $\omega$ an also enhance the steerability, although its impact is less pronounced than that of $\gamma$.\\
\begin{figure}[H]
\begin{center}
\includegraphics[width=18cm,height=8cm]{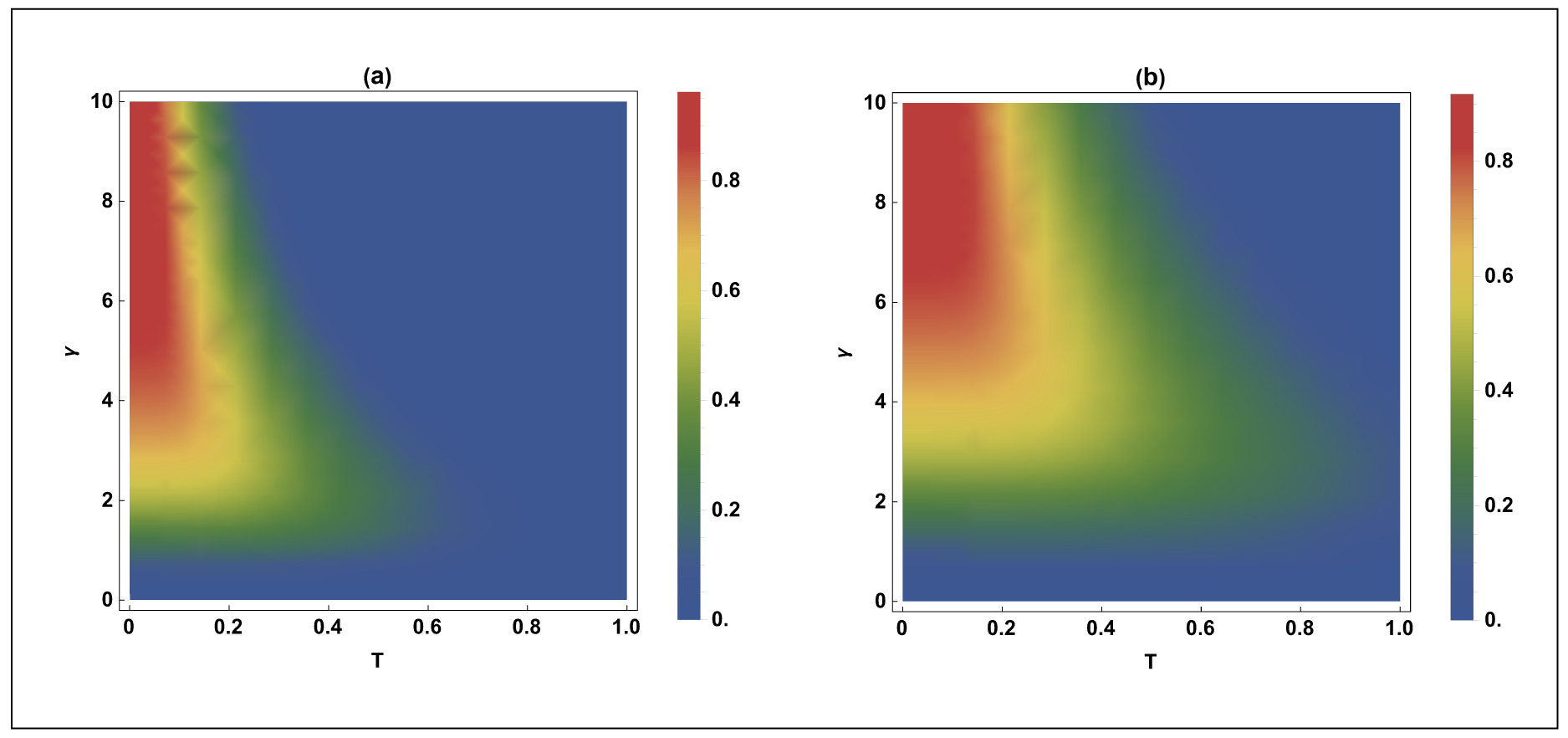}
\end{center}
\caption{Plot of quantum steering in two gravcats versus $\gamma$ and $T$ with (a) $\omega=2$ and (b) $\omega= 3$.}
\label{fig:4}
\end{figure}
The effects of temperature on steerability are clearly visible in Fig. \ref{fig:4}, where we observe steerability as a function of $\gamma$ and $T$. As the temperature increases, thermal effects become predominant, leading to an increase in entropy in the system and a decrease in steerability. Comparing the Fig. \ref{fig:4}(a) and Fig. \ref{fig:4}(b), we note that increasing the excitation energy $\omega$  leads to greater energy availability for quantum processes,  thus enhancing steerability under constant other variables. Moreover, the gravitational potential energy introduces a unique element: strong gravitational interactions can either amplify or dampen quantum effects, resulting in steerability demonstrating non-linear behavior.\\
\begin{figure}[H]
\begin{center}
\includegraphics[width=18cm,height=8cm]{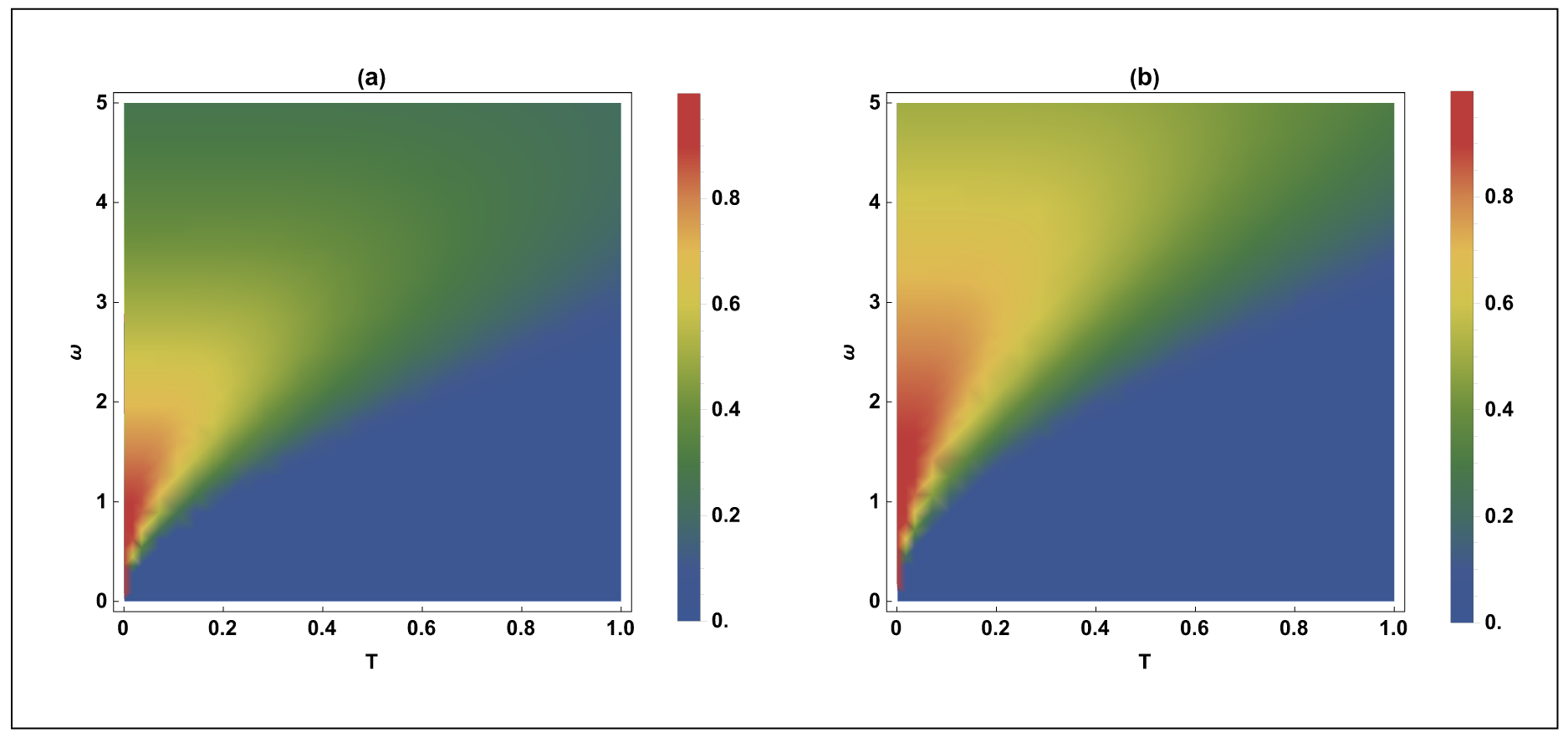}
\end{center}
\caption{Plot of quantum steering in two gravcats versus $\omega$ and $T$ with (a) $\gamma=3$ and (b) $\gamma=5$.}
\label{fig:5}
\end{figure}
The previously described characteristics of steerability can be confirmed by analyzing Fig.\ref{fig:5}, as the system temperature increases, thermal fluctuations become more significant. This generally leads to increased noise, which reduces the quality of quantum information processing and thus decreases the system's steerability, this phenomenon is evident in Fig. \ref{fig:5}(a) and Fig. \ref{fig:5}(b). Regarding the excitation energy $\omega$,  we observe nonlinear behavior in steerability. At higher energy levels, there is more energy available for quantum processes, which can enhance the system's steerability for manipulating and processing quantum information. However, at extremely high values of $\omega$, other factors such as increased thermal effects may counteract this trend. Comparing Fig.\ref{fig:5}(a) and Fig. \ref{fig:5}(b), we notice an interaction between gravitational effects and steerability. This suggests that the strength of gravitational interactions can influence how the quantum system can be used for quantum communication, which could have significant implications for the design and optimization of quantum systems in environments where gravitational effects are significant.\\
\begin{figure}[H]
\begin{minipage}{0.5\linewidth}
\includegraphics[scale=0.45]{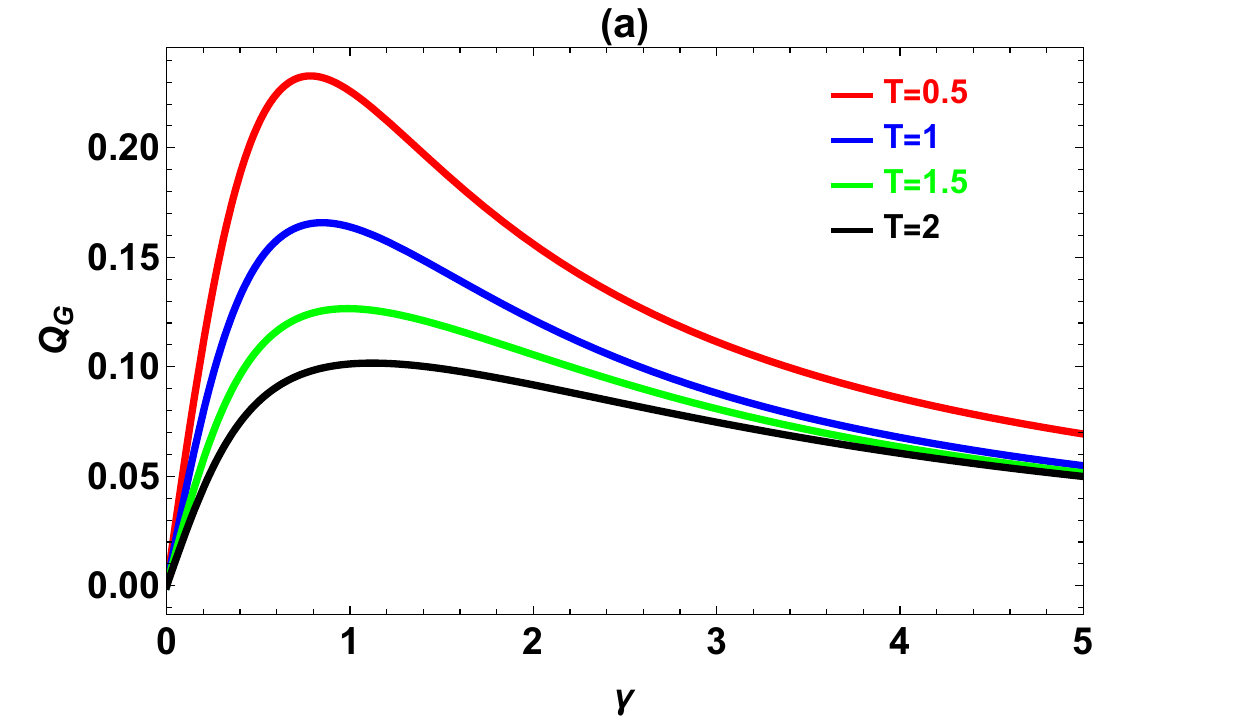}
\end{minipage}\hfill
\begin{minipage}{0.5\linewidth}
\includegraphics[scale=0.45]{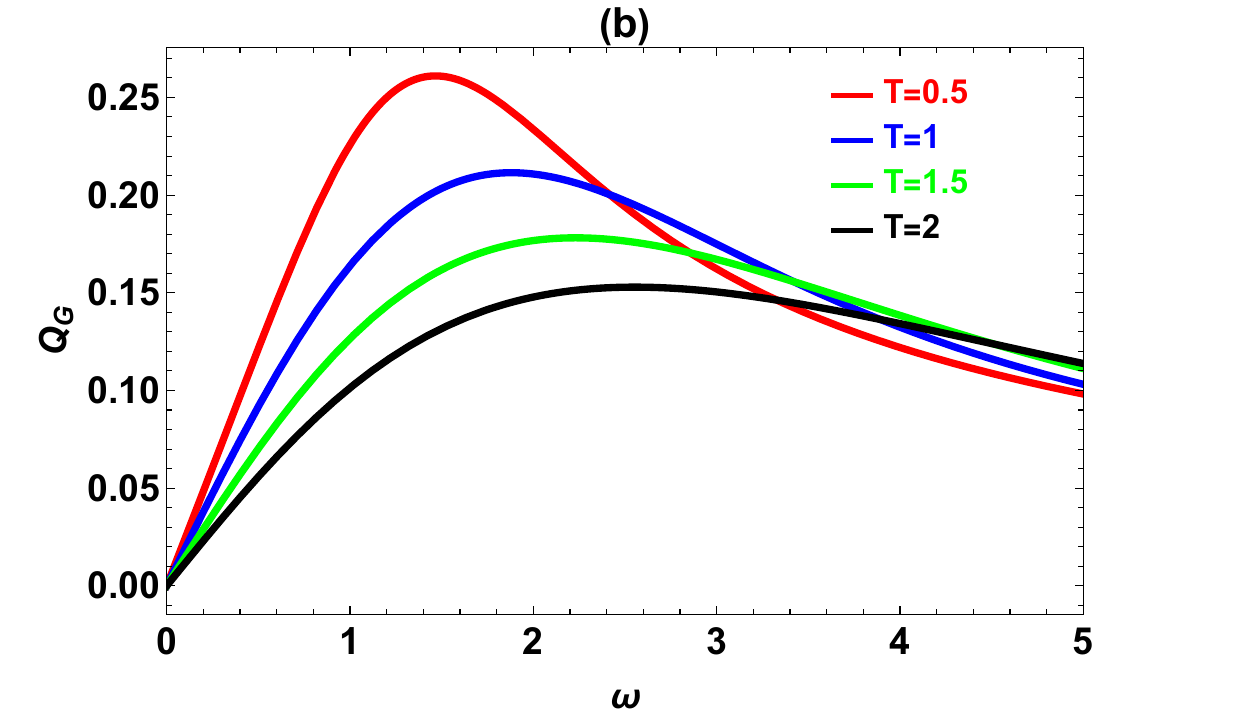}
\end{minipage}
\caption{Plot of geometric quantum discord $Q_{G}$ versus (a) $\gamma$ and (b) $\omega$ for various values of temperature $T$.}
\label{fig:6}
\end{figure} 
In Fig.\ref{fig:6}, we plot geometric quantum discord $Q_{G}$ as a function of potential energy $\gamma$ (Fig.\ref{fig:6}(a)) and excitation energy $\omega$ (Fig.\ref{fig:6}(b)) for various values of temperature $T$. We observe that as we increase the parameters $\gamma$ or $\omega$, quantum discord $Q_{G}$ rapidly increases to reach its maximum and then decreases to reach a steady state value. It is also noteworthy that the maximum value of quantum discord $Q_{G}$ occurs at low temperature and lower potential energy $\gamma$, as depicted in Fig.\ref{fig:6}(a), while at higher excitation energies, the maximum occurs at a higher energy level, as seen in Fig.\ref{fig:6}(b). This suggests a complex relationship between temperature $T$, potential energy, excitation energy, and the measurement of geometric quantum discord.\\
\begin{figure}[H]
\begin{minipage}{0.5\linewidth}
\includegraphics[scale=0.45]{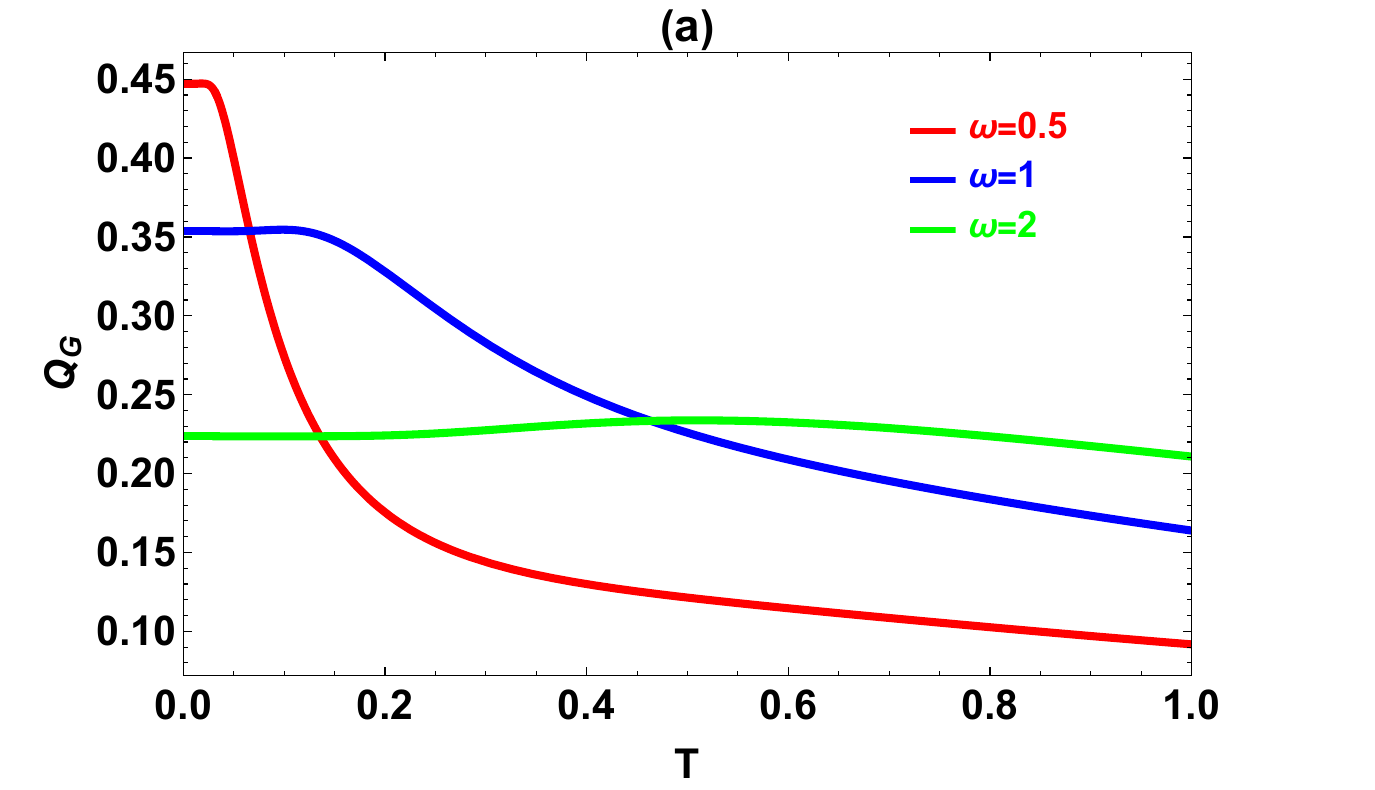}
\end{minipage}\hfill
\begin{minipage}{0.5\linewidth}
\includegraphics[scale=0.45]{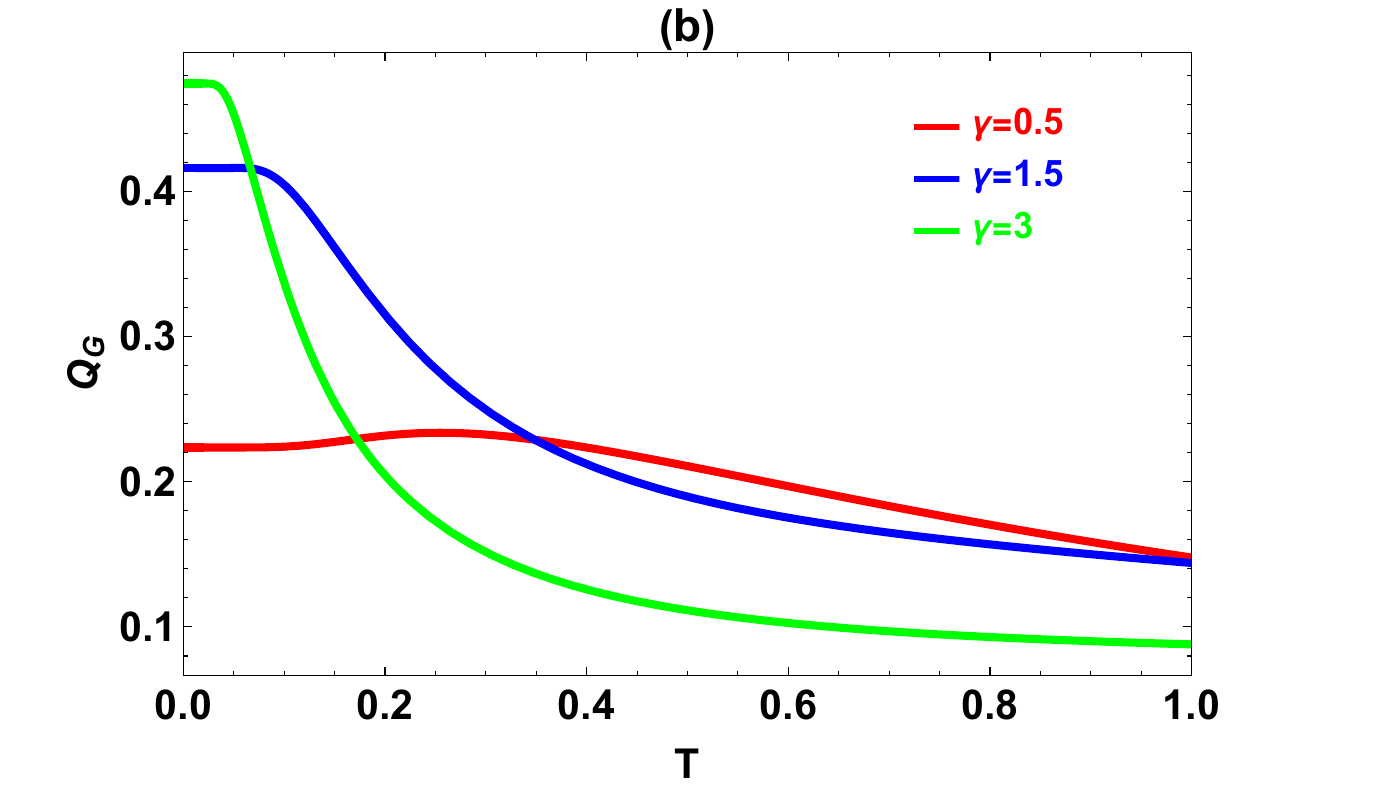}
\end{minipage}
\caption{Plot of geometric quantum discord $Q_{G}$ versus temperature $T$ for different values of (a) $\gamma =1$ and (b) $\omega =1$.}
\label{fig:7}
\end{figure}
In Fig.\ref{fig:7}(a), we depict the quantum discord $Q_{G}$ as a function of temperature $T$ for various values of the potential energy. Meanwhile, in Fig.\ref{fig:7}(b), we illustrate the quantum discord $Q_{G}$ as a function of $T$ for different values of $\omega$. As the temperature increases, the quantum discord decreases after reaching a certain threshold, which is $T=0.02$ for specific values of $\omega=0.5$ and $\gamma=1$, as depicted in Fig.\ref{fig:7}(a). However, for a different value of $\omega$, for example, $\omega=2$, the GQD decreases more smoothly and steadily around $0.2$. At lower temperatures, increasing $\omega$ can lead to an increase in the $Q_{G}$. However, at higher temperatures, this relationship may no longer hold true, and it also depends on the values of $\gamma$.

\begin{figure}[H]
\begin{minipage}{0.33\linewidth}
\includegraphics[scale=0.3]{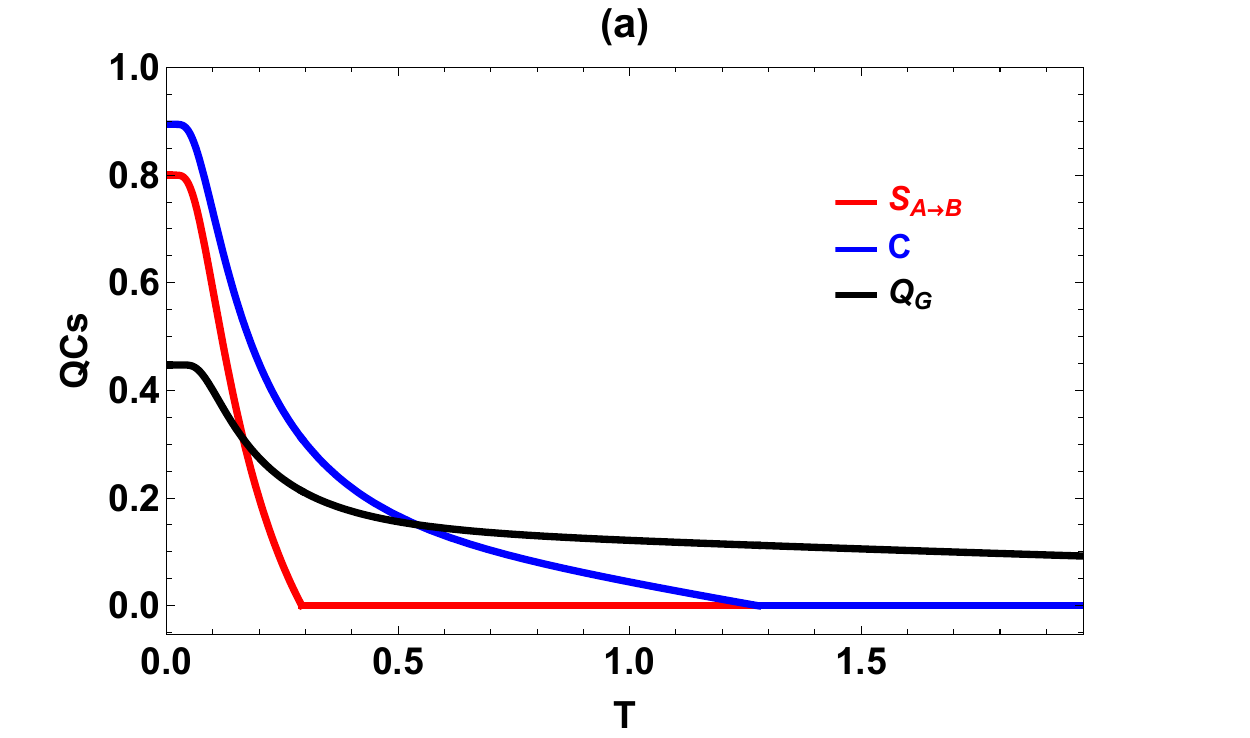}
\end{minipage}\hfill
\begin{minipage}{0.33\linewidth}
\includegraphics[scale=0.3]{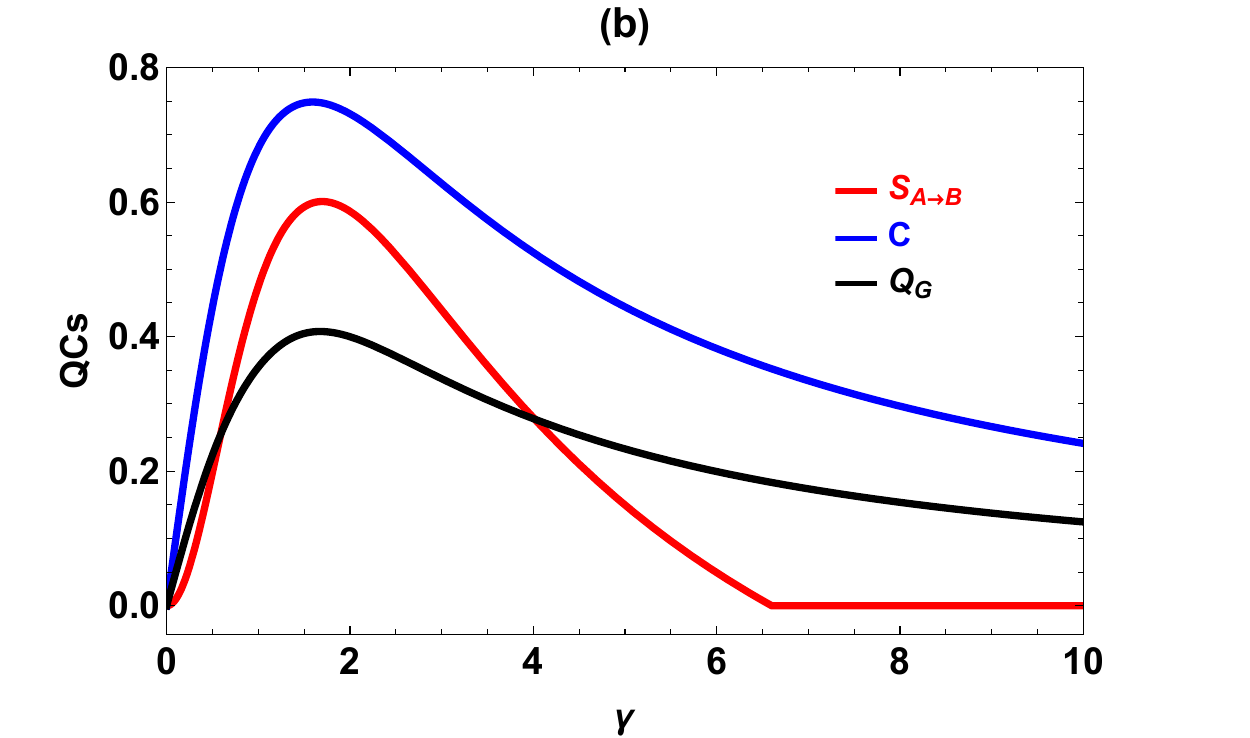}
\end{minipage}\hfill
\begin{minipage}{0.33\linewidth}
\includegraphics[scale=0.3]{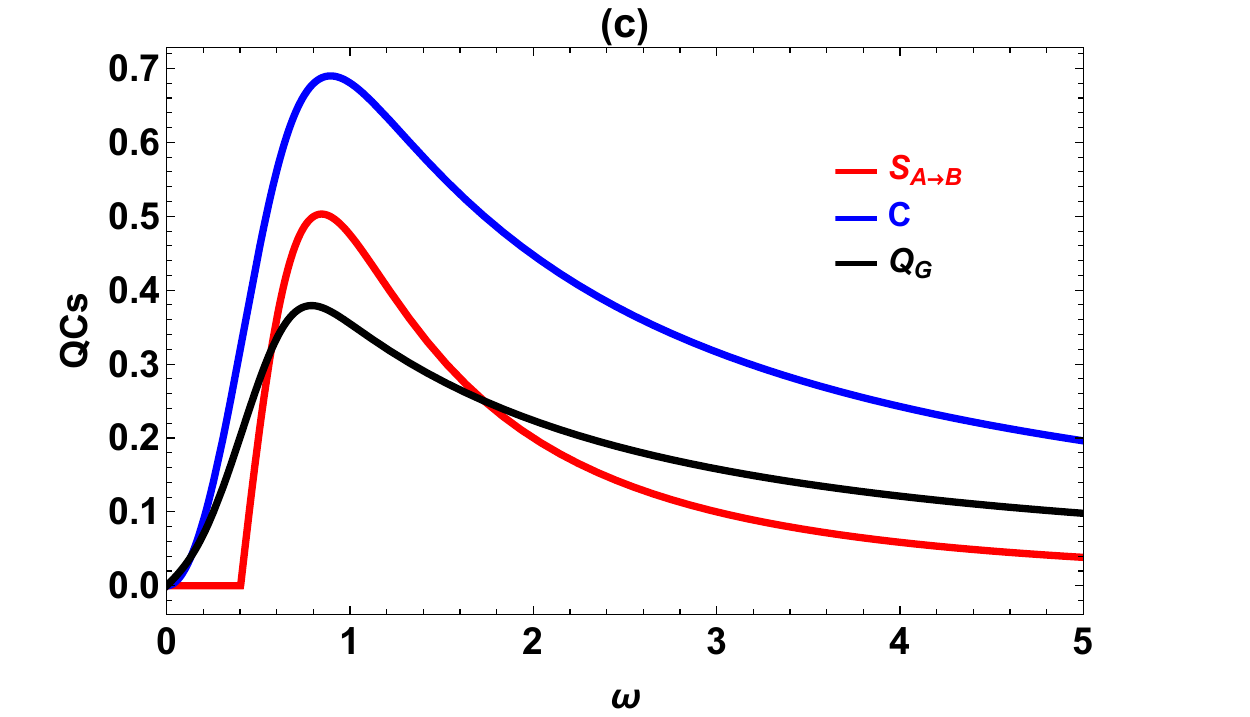}
\end{minipage}
\caption{Plot of the quantum steering $S_{A\rightarrow B}$, concurrence $C$ and geometric quantum discord $Q_{G}$ (a) as a function of temperature with $\gamma =2$ and $\omega =1$, (b) as a fuction of $\gamma$ with $T=0.1$ and $\omega =1$, (c) as a function of $\omega$ with $T=0.1$ and $\gamma =1$.}
\label{fig:8}
\end{figure}

In Fig.\ref{fig:8}(a), we investigate quantum correlations such as quantum steering $S_{A\rightarrow B}$, concurrence $C$ and quantum discord $Q_{G}$ as a function of thermal noise. This figure illustrates the hierarchy of quantum correlations, revealing how they evolve with temperature. When $0.28 < T < 1.2$, quantum discord and concurrence remain significant, while the steering $S_{A\rightarrow B}$ is zero. This suggests that the two gravcats are entangled ($C>0$). Furthermore, it can be observed that the directed state can be entangled even when the entangled state is not necessarily steered. It is also noted that the steering $S_{A\rightarrow B}$ is bounded by the concurrence $C$. Additionally, when $T>1.2$, quantum correlations beyond entanglement manifest between the two gravcats, as quantum discord persists despite the absence of concurrence.

In Fig. \ref{fig:8}, we plotted quantum steering $S_{A\rightarrow B}$, concurrence $C$ and quantum discord $Q_{G}$ as functions of potential energy $\gamma$ (Fig.\ref{fig:8}(b)) and as functions of excitation energy $\omega$ (Fig.\ref{fig:8}(c)). We observe that quantum steering $S_{A\rightarrow B}$, quantum discord $Q_{G}$ and concurrence $C$ exhibit similar behaviors. When $\gamma<6.6$ (or $\omega<0.4$) in Fig.\ref{fig:8}(b) (or in Fig.\ref{fig:8}(c), quantum steering is zero, while concurrence is positive. This implies that a steerable state must be entangled, even though not all entangled states are necessarily steerable, as illustrated in Fig.\ref{fig:7}. However, within the region where entanglement is absent, the quantum discord of the system is still considerably greater than zero, ensuring the presence of correlated quantum states even when the system is in a separable state.

\section{Conclusion} \label{sec:5}

In Summery, the hierarchy of quantum correlations between two gravitational cats states. We have studied the thermal effects generated by a thermal bath on the quantum correlations induced by the gravitational interaction of a system of two massive particles confined in two distinct double-well potentials. We have discussed and compared different quantum correlations (quantum steering , concurrence, and quantum discord) under different parameters. We have shown that quantum entanglement remains more persistent than quantum steering, i.e., the directed modes are strictly entangled, but the entangled modes are not always directed. Besides, we have studied quantum correlations beyond entanglement via GQD. At higher temperatures, particles in a system tend to interact more with their thermal environment, which can lead to a phenomenon called thermal decoherence. Also, the influence of the potential energy and excitation energy  on the quantum correlations have been considered.


{\small\bibliographystyle{unsrtnat}
\bibliography{sample}}

\begin{thebibliography}{66}
\expandafter\ifx\csname natexlab\endcsname\relax\def\natexlab#1{#1}\fi
\expandafter\ifx\csname bibnamefont\endcsname\relax
  \def\bibnamefont#1{#1}\fi
\expandafter\ifx\csname bibfnamefont\endcsname\relax
  \def\bibfnamefont#1{#1}\fi
\expandafter\ifx\csname citenamefont\endcsname\relax
  \def\citenamefont#1{#1}\fi
\expandafter\ifx\csname url\endcsname\relax
  \def\url#1{\texttt{#1}}\fi
\expandafter\ifx\csname urlprefix\endcsname\relax\def\urlprefix{URL }\fi
\providecommand{\bibinfo}[2]{#2}
\providecommand{\eprint}[2][]{\url{#2}}

\bibitem[{\citenamefont{Cirac et~al.}(1998)\citenamefont{Cirac, Lewenstein,
  Molmer, and Zoller}}]{1}
\bibinfo{author}{\bibfnamefont{J.}~\bibnamefont{Cirac}},
  \bibinfo{author}{\bibfnamefont{K.}~\bibnamefont{Lewenstein}},
  \bibinfo{author}{\bibfnamefont{K.}~\bibnamefont{Molmer}}, \bibnamefont{and}
  \bibinfo{author}{\bibfnamefont{P.}~\bibnamefont{Zoller}},
  \bibinfo{journal}{Physical Review A} \textbf{\bibinfo{volume}{57(2)}},
  \bibinfo{pages}{1202} (\bibinfo{year}{1998}).

\bibitem[{\citenamefont{Gerry and Knight}(1998)}]{2}
\bibinfo{author}{\bibfnamefont{C.}~\bibnamefont{Gerry}} \bibnamefont{and}
  \bibinfo{author}{\bibfnamefont{P.}~\bibnamefont{Knight}},
  \bibinfo{journal}{American Journal of Physics}
  \textbf{\bibinfo{volume}{65(10)}}, \bibinfo{pages}{964}
  (\bibinfo{year}{1998}).

\bibitem[{\citenamefont{Horodecki et~al.}(2009)\citenamefont{Horodecki,
  Horodecki, Horodecki, and Horodecki}}]{3}
\bibinfo{author}{\bibfnamefont{R.}~\bibnamefont{Horodecki}},
  \bibinfo{author}{\bibfnamefont{P.}~\bibnamefont{Horodecki}},
  \bibinfo{author}{\bibfnamefont{M.}~\bibnamefont{Horodecki}},
  \bibnamefont{and}
  \bibinfo{author}{\bibfnamefont{K.}~\bibnamefont{Horodecki}},
  \bibinfo{journal}{Reviews of modern physics}
  \textbf{\bibinfo{volume}{81(2)}}, \bibinfo{pages}{865}
  (\bibinfo{year}{2009}).

\bibitem[{\citenamefont{Zyczkowski et~al.}(2001)\citenamefont{Zyczkowski,
  Horodecki, Horodecki, and Horodecki}}]{4}
\bibinfo{author}{\bibfnamefont{K.}~\bibnamefont{Zyczkowski}},
  \bibinfo{author}{\bibfnamefont{P.}~\bibnamefont{Horodecki}},
  \bibinfo{author}{\bibfnamefont{M.}~\bibnamefont{Horodecki}},
  \bibnamefont{and}
  \bibinfo{author}{\bibfnamefont{R.}~\bibnamefont{Horodecki}},
  \bibinfo{journal}{Physical Review A} \textbf{\bibinfo{volume}{65(1)}},
  \bibinfo{pages}{012101} (\bibinfo{year}{2001}).

\bibitem[{\citenamefont{Brito et~al.}(2018)\citenamefont{Brito, Amaral, and
  Chaves}}]{5}
\bibinfo{author}{\bibfnamefont{S.}~\bibnamefont{Brito}},
  \bibinfo{author}{\bibfnamefont{B.}~\bibnamefont{Amaral}}, \bibnamefont{and}
  \bibinfo{author}{\bibfnamefont{R.}~\bibnamefont{Chaves}},
  \bibinfo{journal}{Physical Review A} \textbf{\bibinfo{volume}{97(2)}},
  \bibinfo{pages}{022111} (\bibinfo{year}{2018}).

\bibitem[{\citenamefont{Wiseman et~al.}(2007)\citenamefont{Wiseman, Jones, and
  Doherty}}]{8}
\bibinfo{author}{\bibfnamefont{H.}~\bibnamefont{Wiseman}},
  \bibinfo{author}{\bibfnamefont{S.}~\bibnamefont{Jones}}, \bibnamefont{and}
  \bibinfo{author}{\bibfnamefont{A.}~\bibnamefont{Doherty}},
  \bibinfo{journal}{Physical review letters} \textbf{\bibinfo{volume}{98(14)}},
  \bibinfo{pages}{140402} (\bibinfo{year}{2007}).

\bibitem[{\citenamefont{Uola et~al.}(2020)\citenamefont{Uola, Costa, Nguyen,
  and Günter}}]{9}
\bibinfo{author}{\bibfnamefont{R.}~\bibnamefont{Uola}},
  \bibinfo{author}{\bibfnamefont{A.}~\bibnamefont{Costa}},
  \bibinfo{author}{\bibfnamefont{H.}~\bibnamefont{Nguyen}}, \bibnamefont{and}
  \bibinfo{author}{\bibfnamefont{O.}~\bibnamefont{Günter}},
  \bibinfo{journal}{Reviews of Modern Physics}
  \textbf{\bibinfo{volume}{92(1)}}, \bibinfo{pages}{015001}
  (\bibinfo{year}{2020}).

\bibitem[{\citenamefont{Ollivier and Zurek}(2001{\natexlab{a}})}]{6}
\bibinfo{author}{\bibfnamefont{H.}~\bibnamefont{Ollivier}} \bibnamefont{and}
  \bibinfo{author}{\bibfnamefont{W.}~\bibnamefont{Zurek}},
  \bibinfo{journal}{Physical review letters} \textbf{\bibinfo{volume}{017901}},
  \bibinfo{pages}{88(1)} (\bibinfo{year}{2001}{\natexlab{a}}).

\bibitem[{\citenamefont{Luo and Fu}(2010)}]{7}
\bibinfo{author}{\bibfnamefont{S.}~\bibnamefont{Luo}} \bibnamefont{and}
  \bibinfo{author}{\bibfnamefont{S.}~\bibnamefont{Fu}},
  \bibinfo{journal}{Physical Review A} \textbf{\bibinfo{volume}{82(3)}},
  \bibinfo{pages}{034302} (\bibinfo{year}{2010}).

\bibitem[{\citenamefont{Laflorencie}(2016)}]{10}
\bibinfo{author}{\bibfnamefont{N.}~\bibnamefont{Laflorencie}},
  \bibinfo{journal}{Physics Reports} \textbf{\bibinfo{volume}{646}},
  \bibinfo{pages}{1} (\bibinfo{year}{2016}).

\bibitem[{\citenamefont{O’Brien et~al.}(2009)\citenamefont{O’Brien,
  Furusawa, and Vučković}}]{11}
\bibinfo{author}{\bibfnamefont{J.-L.} \bibnamefont{O’Brien}},
  \bibinfo{author}{\bibfnamefont{A.}~\bibnamefont{Furusawa}}, \bibnamefont{and}
  \bibinfo{author}{\bibfnamefont{J.}~\bibnamefont{Vučković}},
  \bibinfo{journal}{Nature photonics} \textbf{\bibinfo{volume}{3(12)}},
  \bibinfo{pages}{687} (\bibinfo{year}{2009}).

\bibitem[{\citenamefont{Ladd et~al.}(2010)\citenamefont{Ladd, Jelezko,
  Laflamme, Nakamura, Monroe, and O’Brien}}]{12}
\bibinfo{author}{\bibfnamefont{T.}~\bibnamefont{Ladd}},
  \bibinfo{author}{\bibfnamefont{F.}~\bibnamefont{Jelezko}},
  \bibinfo{author}{\bibfnamefont{R.}~\bibnamefont{Laflamme}},
  \bibinfo{author}{\bibfnamefont{Y.}~\bibnamefont{Nakamura}},
  \bibinfo{author}{\bibfnamefont{C.}~\bibnamefont{Monroe}}, \bibnamefont{and}
  \bibinfo{author}{\bibfnamefont{J.}~\bibnamefont{O’Brien}},
  \bibinfo{journal}{nature} \textbf{\bibinfo{volume}{464(7285)}},
  \bibinfo{pages}{45} (\bibinfo{year}{2010}).

\bibitem[{\citenamefont{Jones and Mosca}(1998)}]{13}
\bibinfo{author}{\bibfnamefont{J.}~\bibnamefont{Jones}} \bibnamefont{and}
  \bibinfo{author}{\bibfnamefont{M.}~\bibnamefont{Mosca}},
  \bibinfo{journal}{The Journal of chemical physics}
  \textbf{\bibinfo{volume}{109(5)}}, \bibinfo{pages}{1648}
  (\bibinfo{year}{1998}).

\bibitem[{\citenamefont{Wang et~al.}(2020)\citenamefont{Wang, Sciarrino, Laing,
  and Thompson}}]{14}
\bibinfo{author}{\bibfnamefont{J.}~\bibnamefont{Wang}},
  \bibinfo{author}{\bibfnamefont{F.}~\bibnamefont{Sciarrino}},
  \bibinfo{author}{\bibfnamefont{A.}~\bibnamefont{Laing}}, \bibnamefont{and}
  \bibinfo{author}{\bibfnamefont{M.}~\bibnamefont{Thompson}},
  \bibinfo{journal}{Nature Photonics} \textbf{\bibinfo{volume}{14(5))}},
  \bibinfo{pages}{273} (\bibinfo{year}{2020}).

\bibitem[{\citenamefont{Auffèves}(2022)}]{15}
\bibinfo{author}{\bibfnamefont{A.}~\bibnamefont{Auffèves}},
  \bibinfo{journal}{PRX Quantum} \textbf{\bibinfo{volume}{3(2)}},
  \bibinfo{pages}{020101} (\bibinfo{year}{2022}).

\bibitem[{\citenamefont{WEINBERG}(1989)}]{16}
\bibinfo{author}{\bibfnamefont{S.}~\bibnamefont{WEINBERG}},
  \bibinfo{journal}{Annals of Physics} \textbf{\bibinfo{volume}{194(2)}},
  \bibinfo{pages}{336} (\bibinfo{year}{1989}).

\bibitem[{\citenamefont{Omar et~al.}(2006)\citenamefont{Omar, Paunković,
  Sheridan, and Bose}}]{17}
\bibinfo{author}{\bibfnamefont{Y.}~\bibnamefont{Omar}},
  \bibinfo{author}{\bibfnamefont{N.}~\bibnamefont{Paunković}},
  \bibinfo{author}{\bibfnamefont{L.}~\bibnamefont{Sheridan}}, \bibnamefont{and}
  \bibinfo{author}{\bibfnamefont{S.}~\bibnamefont{Bose}},
  \bibinfo{journal}{Physical Review A} \textbf{\bibinfo{volume}{74(4)}},
  \bibinfo{pages}{042304} (\bibinfo{year}{2006}).

\bibitem[{\citenamefont{Ma et~al.}(2009)\citenamefont{Ma, Qarry, Kofler,
  Jennewein, and Zeilinger1.A}}]{18}
\bibinfo{author}{\bibfnamefont{X.}~\bibnamefont{Ma}},
  \bibinfo{author}{\bibfnamefont{A.}~\bibnamefont{Qarry}},
  \bibinfo{author}{\bibfnamefont{J.}~\bibnamefont{Kofler}},
  \bibinfo{author}{\bibfnamefont{T.}~\bibnamefont{Jennewein}},
  \bibnamefont{and} \bibinfo{author}{\bibnamefont{Zeilinger1.A}},
  \bibinfo{journal}{Physical Review A} \textbf{\bibinfo{volume}{79(4)}},
  \bibinfo{pages}{042101} (\bibinfo{year}{2009}).

\bibitem[{\citenamefont{Das et~al.}(2019)\citenamefont{Das, Sasmal, and
  Roy}}]{19}
\bibinfo{author}{\bibfnamefont{D.}~\bibnamefont{Das}},
  \bibinfo{author}{\bibfnamefont{S.}~\bibnamefont{Sasmal}}, \bibnamefont{and}
  \bibinfo{author}{\bibfnamefont{S.}~\bibnamefont{Roy}},
  \bibinfo{journal}{Physical Review A} \textbf{\bibinfo{volume}{99(5)}},
  \bibinfo{pages}{052109} (\bibinfo{year}{2019}).

\bibitem[{\citenamefont{Schrodinger}(2018)}]{20}
\bibinfo{author}{\bibfnamefont{E.}~\bibnamefont{Schrodinger}},
  \bibinfo{journal}{Mathematical Proceedings of the Cambridge Philosophical
  Society} \textbf{\bibinfo{volume}{31(4)}}, \bibinfo{pages}{555}
  (\bibinfo{year}{2018}).

\bibitem[{\citenamefont{Brunner et~al.}(2023)\citenamefont{Brunner, Cavalcanti,
  Pironio, Scarani, and Wehner}}]{21}
\bibinfo{author}{\bibfnamefont{N.}~\bibnamefont{Brunner}},
  \bibinfo{author}{\bibfnamefont{D.}~\bibnamefont{Cavalcanti}},
  \bibinfo{author}{\bibfnamefont{S.}~\bibnamefont{Pironio}},
  \bibinfo{author}{\bibfnamefont{V.}~\bibnamefont{Scarani}}, \bibnamefont{and}
  \bibinfo{author}{\bibfnamefont{S.}~\bibnamefont{Wehner}},
  \bibinfo{journal}{Reviews of modern physics}
  \textbf{\bibinfo{volume}{86(2)}}, \bibinfo{pages}{419}
  (\bibinfo{year}{2023}).

\bibitem[{\citenamefont{Hu et~al.}(2018{\natexlab{a}})\citenamefont{Hu, Hu,
  Wang, Peng, Zhang, , and Fan}}]{72}
\bibinfo{author}{\bibfnamefont{M.}~\bibnamefont{Hu}},
  \bibinfo{author}{\bibfnamefont{X.}~\bibnamefont{Hu}},
  \bibinfo{author}{\bibfnamefont{J.}~\bibnamefont{Wang}},
  \bibinfo{author}{\bibfnamefont{Y.}~\bibnamefont{Peng}},
  \bibinfo{author}{\bibfnamefont{Y.~R.} \bibnamefont{Zhang}}, ,
  \bibnamefont{and} \bibinfo{author}{\bibfnamefont{H.}~\bibnamefont{Fan}},
  \bibinfo{journal}{Physics Reports} \textbf{\bibinfo{volume}{762}},
  \bibinfo{pages}{1} (\bibinfo{year}{2018}{\natexlab{a}}).

\bibitem[{\citenamefont{Cruz et~al.}(2022)\citenamefont{Cruz, Anka, Reis,
  Bachelard, and Santos}}]{73}
\bibinfo{author}{\bibfnamefont{C.}~\bibnamefont{Cruz}},
  \bibinfo{author}{\bibfnamefont{M.}~\bibnamefont{Anka}},
  \bibinfo{author}{\bibfnamefont{M.}~\bibnamefont{Reis}},
  \bibinfo{author}{\bibfnamefont{R.}~\bibnamefont{Bachelard}},
  \bibnamefont{and} \bibinfo{author}{\bibfnamefont{A.}~\bibnamefont{Santos}},
  \bibinfo{journal}{Quantum Science and Technology}
  \textbf{\bibinfo{volume}{7(2)}}, \bibinfo{pages}{025020}
  (\bibinfo{year}{2022}).

\bibitem[{\citenamefont{Khedif et~al.}(2022{\natexlab{a}})\citenamefont{Khedif,
  Haddadi, Daoud, Dolatkhah, and Pourkarimi}}]{74}
\bibinfo{author}{\bibfnamefont{Y.}~\bibnamefont{Khedif}},
  \bibinfo{author}{\bibfnamefont{S.}~\bibnamefont{Haddadi}},
  \bibinfo{author}{\bibfnamefont{M.}~\bibnamefont{Daoud}},
  \bibinfo{author}{\bibfnamefont{H.}~\bibnamefont{Dolatkhah}},
  \bibnamefont{and}
  \bibinfo{author}{\bibfnamefont{M.}~\bibnamefont{Pourkarimi}},
  \bibinfo{journal}{Quantum Information Processing}
  \textbf{\bibinfo{volume}{21(7)}}, \bibinfo{pages}{235}
  (\bibinfo{year}{2022}{\natexlab{a}}).

\bibitem[{\citenamefont{Cruz}(2017)}]{75}
\bibinfo{author}{\bibfnamefont{C.}~\bibnamefont{Cruz}},
  \bibinfo{journal}{International Journal of Quantum Information}
  \textbf{\bibinfo{volume}{15(05)}}, \bibinfo{pages}{1750031}
  (\bibinfo{year}{2017}).

\bibitem[{\citenamefont{Deng et~al.}(2017)\citenamefont{Deng, Ren, and
  Li}}]{44}
\bibinfo{author}{\bibfnamefont{F.}~\bibnamefont{Deng}},
  \bibinfo{author}{\bibfnamefont{B.}~\bibnamefont{Ren}}, \bibnamefont{and}
  \bibinfo{author}{\bibfnamefont{X.}~\bibnamefont{Li}},
  \bibinfo{journal}{Science bulletin} \textbf{\bibinfo{volume}{62(1)}},
  \bibinfo{pages}{46} (\bibinfo{year}{2017}).

\bibitem[{\citenamefont{Lo et~al.}(2000)\citenamefont{Lo, Ren, and Li}}]{45}
\bibinfo{author}{\bibfnamefont{H.}~\bibnamefont{Lo}},
  \bibinfo{author}{\bibfnamefont{B.}~\bibnamefont{Ren}}, \bibnamefont{and}
  \bibinfo{author}{\bibfnamefont{X.}~\bibnamefont{Li}},
  \bibinfo{journal}{Physical Review A} \textbf{\bibinfo{volume}{62(1)}},
  \bibinfo{pages}{012313} (\bibinfo{year}{2000}).

\bibitem[{\citenamefont{Glauber}(1963)}]{46}
\bibinfo{author}{\bibfnamefont{R.}~\bibnamefont{Glauber}},
  \bibinfo{journal}{Physical Review} \textbf{\bibinfo{volume}{131(6)}},
  \bibinfo{pages}{2766} (\bibinfo{year}{1963}).

\bibitem[{\citenamefont{Sudarshan}(1963)}]{47}
\bibinfo{author}{\bibfnamefont{E.}~\bibnamefont{Sudarshan}},
  \bibinfo{journal}{Physical Review Letters} \textbf{\bibinfo{volume}{10(7)}},
  \bibinfo{pages}{277} (\bibinfo{year}{1963}).

\bibitem[{\citenamefont{Kammerlander and Anders}(2016)}]{48}
\bibinfo{author}{\bibfnamefont{P.}~\bibnamefont{Kammerlander}}
  \bibnamefont{and} \bibinfo{author}{\bibfnamefont{J.}~\bibnamefont{Anders}},
  \bibinfo{journal}{Scientific reports} \textbf{\bibinfo{volume}{6(1)}},
  \bibinfo{pages}{22174} (\bibinfo{year}{2016}).

\bibitem[{\citenamefont{Zhang et~al.}(2022)\citenamefont{Zhang, Wang, Zeng, and
  Wang}}]{49}
\bibinfo{author}{\bibfnamefont{K.}~\bibnamefont{Zhang}},
  \bibinfo{author}{\bibfnamefont{X.}~\bibnamefont{Wang}},
  \bibinfo{author}{\bibfnamefont{Q.}~\bibnamefont{Zeng}}, \bibnamefont{and}
  \bibinfo{author}{\bibfnamefont{J.}~\bibnamefont{Wang}}, \bibinfo{journal}{PRX
  Quantum} \textbf{\bibinfo{volume}{3(3)}}, \bibinfo{pages}{030315}
  (\bibinfo{year}{2022}).

\bibitem[{\citenamefont{Filgueiras et~al.}(2020)\citenamefont{Filgueiras,
  Rojas, and Rojas}}]{50}
\bibinfo{author}{\bibfnamefont{C.}~\bibnamefont{Filgueiras}},
  \bibinfo{author}{\bibfnamefont{O.}~\bibnamefont{Rojas}}, \bibnamefont{and}
  \bibinfo{author}{\bibfnamefont{M.}~\bibnamefont{Rojas}},
  \bibinfo{journal}{Annalen der Physik} \textbf{\bibinfo{volume}{532(8)}},
  \bibinfo{pages}{2000207} (\bibinfo{year}{2020}).

\bibitem[{\citenamefont{Berrada}(2020)}]{51}
\bibinfo{author}{\bibfnamefont{K.}~\bibnamefont{Berrada}},
  \bibinfo{journal}{Physica E: Low-dimensional Systems and Nanostructures}
  \textbf{\bibinfo{volume}{116}}, \bibinfo{pages}{113784}
  (\bibinfo{year}{2020}).

\bibitem[{\citenamefont{Sha et~al.}(2018)\citenamefont{Sha, Wang, Sun, , and
  Hou}}]{52}
\bibinfo{author}{\bibfnamefont{Y.}~\bibnamefont{Sha}},
  \bibinfo{author}{\bibfnamefont{Y.}~\bibnamefont{Wang}},
  \bibinfo{author}{\bibfnamefont{Z.}~\bibnamefont{Sun}}, , \bibnamefont{and}
  \bibinfo{author}{\bibfnamefont{X.}~\bibnamefont{Hou}},
  \bibinfo{journal}{Annals of Physics} \textbf{\bibinfo{volume}{392}},
  \bibinfo{pages}{229} (\bibinfo{year}{2018}).

\bibitem[{\citenamefont{Zad and Rojas}(2021)}]{53}
\bibinfo{author}{\bibfnamefont{H.}~\bibnamefont{Zad}} \bibnamefont{and}
  \bibinfo{author}{\bibfnamefont{M.}~\bibnamefont{Rojas}},
  \bibinfo{journal}{Physica E: Low-dimensional Systems and Nanostructures}
  \textbf{\bibinfo{volume}{126}}, \bibinfo{pages}{114455}
  (\bibinfo{year}{2021}).

\bibitem[{\citenamefont{Ullah et~al.}(2022)\citenamefont{Ullah, Köse, Yagan,
  and Onbaşlı}}]{54}
\bibinfo{author}{\bibfnamefont{K.}~\bibnamefont{Ullah}},
  \bibinfo{author}{\bibfnamefont{E.}~\bibnamefont{Köse}},
  \bibinfo{author}{\bibfnamefont{R.}~\bibnamefont{Yagan}}, \bibnamefont{and}
  \bibinfo{author}{\bibfnamefont{M.}~\bibnamefont{Onbaşlı}},
  \bibinfo{journal}{Physical Review Research} \textbf{\bibinfo{volume}{4(2)}},
  \bibinfo{pages}{023221} (\bibinfo{year}{2022}).

\bibitem[{\citenamefont{Shi et~al.}(2022)\citenamefont{Shi, Ding, Wan, Wang,
  and Yang}}]{55}
\bibinfo{author}{\bibfnamefont{H.}~\bibnamefont{Shi}},
  \bibinfo{author}{\bibfnamefont{S.}~\bibnamefont{Ding}},
  \bibinfo{author}{\bibfnamefont{Q.}~\bibnamefont{Wan}},
  \bibinfo{author}{\bibfnamefont{X.}~\bibnamefont{Wang}}, \bibnamefont{and}
  \bibinfo{author}{\bibfnamefont{W.}~\bibnamefont{Yang}},
  \bibinfo{journal}{Physical Review Letters}
  \textbf{\bibinfo{volume}{129(13)}}, \bibinfo{pages}{130602}
  (\bibinfo{year}{2022}).

\bibitem[{\citenamefont{Ahnefeld et~al.}(2022)\citenamefont{Ahnefeld, Theurer,
  Egloff, Matera, and Plenio}}]{56}
\bibinfo{author}{\bibfnamefont{F.}~\bibnamefont{Ahnefeld}},
  \bibinfo{author}{\bibfnamefont{T.}~\bibnamefont{Theurer}},
  \bibinfo{author}{\bibfnamefont{D.}~\bibnamefont{Egloff}},
  \bibinfo{author}{\bibfnamefont{J.~M.} \bibnamefont{Matera}},
  \bibnamefont{and} \bibinfo{author}{\bibfnamefont{M.}~\bibnamefont{Plenio}},
  \bibinfo{journal}{Review Letters} \textbf{\bibinfo{volume}{129(12)}},
  \bibinfo{pages}{120501} (\bibinfo{year}{2022}).

\bibitem[{\citenamefont{Sk and Panigrahi}(2022)}]{57}
\bibinfo{author}{\bibfnamefont{R.}~\bibnamefont{Sk}} \bibnamefont{and}
  \bibinfo{author}{\bibfnamefont{P.}~\bibnamefont{Panigrahi}},
  \bibinfo{journal}{Physica A: Statistical Mechanics and its Applications}
  \textbf{\bibinfo{volume}{596}}, \bibinfo{pages}{127129}
  (\bibinfo{year}{2022}).

\bibitem[{\citenamefont{Zurek}(2003)}]{29}
\bibinfo{author}{\bibfnamefont{W.~H.} \bibnamefont{Zurek}},
  \bibinfo{journal}{Reviews of Modern Physics}
  \textbf{\bibinfo{volume}{75(3)}}, \bibinfo{pages}{715}
  (\bibinfo{year}{2003}).

\bibitem[{\citenamefont{Hu et~al.}(2018{\natexlab{b}})\citenamefont{Hu, Hu,
  Wang, Peng, Zhang, and Fan}}]{30}
\bibinfo{author}{\bibfnamefont{M.-L.} \bibnamefont{Hu}},
  \bibinfo{author}{\bibfnamefont{X.}~\bibnamefont{Hu}},
  \bibinfo{author}{\bibfnamefont{J.}~\bibnamefont{Wang}},
  \bibinfo{author}{\bibfnamefont{Y.}~\bibnamefont{Peng}},
  \bibinfo{author}{\bibfnamefont{Y.-R.} \bibnamefont{Zhang}}, \bibnamefont{and}
  \bibinfo{author}{\bibfnamefont{H.}~\bibnamefont{Fan}}
  (\bibinfo{year}{2018}{\natexlab{b}}).

\bibitem[{\citenamefont{Rojas and Lobo}(2023)}]{22}
\bibinfo{author}{\bibfnamefont{M.}~\bibnamefont{Rojas}} \bibnamefont{and}
  \bibinfo{author}{\bibfnamefont{P.}~\bibnamefont{Lobo}},
  \bibinfo{journal}{Universe} \textbf{\bibinfo{volume}{9(2)}},
  \bibinfo{pages}{71} (\bibinfo{year}{2023}).

\bibitem[{\citenamefont{Morozov}(1992)}]{23}
\bibinfo{author}{\bibfnamefont{A.}~\bibnamefont{Morozov}},
  \bibinfo{journal}{Soviet Physics Uspekhi} \textbf{\bibinfo{volume}{35(8)}},
  \bibinfo{pages}{671} (\bibinfo{year}{1992}).

\bibitem[{\citenamefont{Rovelli}(2008)}]{24}
\bibinfo{author}{\bibfnamefont{C.}~\bibnamefont{Rovelli}},
  \bibinfo{journal}{Reviews of modern physics} \textbf{\bibinfo{volume}{11}},
  \bibinfo{pages}{1} (\bibinfo{year}{2008}).

\bibitem[{\citenamefont{Chiou}(2014)}]{25}
\bibinfo{author}{\bibfnamefont{D.}~\bibnamefont{Chiou}},
  \bibinfo{journal}{International Journal of Modern Physics D}
  \textbf{\bibinfo{volume}{24(01)}}, \bibinfo{pages}{1530005}
  (\bibinfo{year}{2014}).

\bibitem[{\citenamefont{Wald}(July 2001)}]{26}
\bibinfo{author}{\bibfnamefont{R.}~\bibnamefont{Wald}},
  \bibinfo{journal}{Living reviews in relativity} \textbf{\bibinfo{volume}{4}},
  \bibinfo{pages}{1} (\bibinfo{year}{July 2001}).

\bibitem[{\citenamefont{Townsend}(1997)}]{27}
\bibinfo{author}{\bibfnamefont{P.}~\bibnamefont{Townsend}},
  \bibinfo{journal}{arXiv preprint gr-qc/9707012}  (\bibinfo{year}{1997}).

\bibitem[{\citenamefont{Kawasaki et~al.}(2008)\citenamefont{Kawasaki, Kohri,
  Moroi, and Yotsuyanagi}}]{28}
\bibinfo{author}{\bibfnamefont{M.}~\bibnamefont{Kawasaki}},
  \bibinfo{author}{\bibfnamefont{K.}~\bibnamefont{Kohri}},
  \bibinfo{author}{\bibfnamefont{T.}~\bibnamefont{Moroi}}, \bibnamefont{and}
  \bibinfo{author}{\bibfnamefont{A.}~\bibnamefont{Yotsuyanagi}},
  \bibinfo{journal}{PHYSICAL REVIEW D} \textbf{\bibinfo{volume}{78(6)}},
  \bibinfo{pages}{065011} (\bibinfo{year}{2008}).

\bibitem[{\citenamefont{Anastopoulos and Hu}(2020)}]{31}
\bibinfo{author}{\bibfnamefont{C.}~\bibnamefont{Anastopoulos}}
  \bibnamefont{and} \bibinfo{author}{\bibfnamefont{B.}~\bibnamefont{Hu}},
  \bibinfo{journal}{Classical and Quantum Gravity}
  \textbf{\bibinfo{volume}{37(23)}}, \bibinfo{pages}{235012}
  (\bibinfo{year}{2020}).

\bibitem[{\citenamefont{Wu et~al.}(2024)\citenamefont{Wu, Li, Teng, Huang, and
  Lu}}]{32}
\bibinfo{author}{\bibfnamefont{S.}~\bibnamefont{Wu}},
  \bibinfo{author}{\bibfnamefont{X.}~\bibnamefont{Li},
  \bibfnamefont{J.X.and~Jiang}},
  \bibinfo{author}{\bibfnamefont{X.}~\bibnamefont{Teng}},
  \bibinfo{author}{\bibfnamefont{X.}~\bibnamefont{Huang}}, \bibnamefont{and}
  \bibinfo{author}{\bibfnamefont{J.}~\bibnamefont{Lu}}, \bibinfo{journal}{The
  European Physical Journal} \textbf{\bibinfo{volume}{84}},
  \bibinfo{pages}{161} (\bibinfo{year}{2024}).

\bibitem[{\citenamefont{Chen et~al.}((2018))\citenamefont{Chen, Ren, Ye, and
  Chen}}]{33}
\bibinfo{author}{\bibfnamefont{C.}~\bibnamefont{Chen}},
  \bibinfo{author}{\bibfnamefont{C.}~\bibnamefont{Ren}},
  \bibinfo{author}{\bibfnamefont{X.}~\bibnamefont{Ye}}, \bibnamefont{and}
  \bibinfo{author}{\bibfnamefont{X.-J.} \bibnamefont{Chen}},
  \bibinfo{journal}{arXiv preprint arXiv:1810.10234}  (\bibinfo{year}{(2018)}).

\bibitem[{\citenamefont{Zhang1 and Wang1}(2021)}]{34}
\bibinfo{author}{\bibfnamefont{K.}~\bibnamefont{Zhang1}} \bibnamefont{and}
  \bibinfo{author}{\bibfnamefont{J.}~\bibnamefont{Wang1}},
  \bibinfo{journal}{Physical Review A} \textbf{\bibinfo{volume}{104(4)}},
  \bibinfo{pages}{042404} (\bibinfo{year}{2021}).

\bibitem[{\citenamefont{Galindo and Martin-Delgado}(2002)}]{35}
\bibinfo{author}{\bibfnamefont{A.}~\bibnamefont{Galindo}} \bibnamefont{and}
  \bibinfo{author}{\bibfnamefont{M.}~\bibnamefont{Martin-Delgado}},
  \bibinfo{journal}{Reviews of Modern Physics}
  \textbf{\bibinfo{volume}{74(2)}}, \bibinfo{pages}{347}
  (\bibinfo{year}{2002}).

\bibitem[{\citenamefont{Chen et~al.}(2018)\citenamefont{Chen, Ren, Ye, and
  Chen}}]{36}
\bibinfo{author}{\bibfnamefont{C.}~\bibnamefont{Chen}},
  \bibinfo{author}{\bibfnamefont{C.}~\bibnamefont{Ren}},
  \bibinfo{author}{\bibfnamefont{X.-J.} \bibnamefont{Ye}}, \bibnamefont{and}
  \bibinfo{author}{\bibfnamefont{J.-L.} \bibnamefont{Chen}},
  \bibinfo{journal}{Physical Review A} \textbf{\bibinfo{volume}{98(5)}},
  \bibinfo{pages}{052114} (\bibinfo{year}{2018}).

\bibitem[{\citenamefont{Wootters}(1998)}]{37}
\bibinfo{author}{\bibfnamefont{W.}~\bibnamefont{Wootters}},
  \bibinfo{journal}{Physical Review Letters} \textbf{\bibinfo{volume}{80}},
  \bibinfo{pages}{2245} (\bibinfo{year}{1998}).

\bibitem[{\citenamefont{Hill and Wootters}(1997)}]{42}
\bibinfo{author}{\bibfnamefont{S.}~\bibnamefont{Hill}} \bibnamefont{and}
  \bibinfo{author}{\bibfnamefont{W.}~\bibnamefont{Wootters}},
  \bibinfo{journal}{Physical review letters} \textbf{\bibinfo{volume}{78(26)}},
  \bibinfo{pages}{5022} (\bibinfo{year}{1997}).

\bibitem[{\citenamefont{Ollivier and Zurek}(2001{\natexlab{b}})}]{62}
\bibinfo{author}{\bibfnamefont{H.}~\bibnamefont{Ollivier}} \bibnamefont{and}
  \bibinfo{author}{\bibfnamefont{W.~H.} \bibnamefont{Zurek}},
  \bibinfo{journal}{Physical review letters} \textbf{\bibinfo{volume}{88(1)}},
  \bibinfo{pages}{017901} (\bibinfo{year}{2001}{\natexlab{b}}).

\bibitem[{\citenamefont{Paula et~al.}(2013)\citenamefont{Paula, Oliveira, and
  Sarandy}}]{63}
\bibinfo{author}{\bibfnamefont{F.~M.} \bibnamefont{Paula}},
  \bibinfo{author}{\bibfnamefont{T.~R.} \bibnamefont{Oliveira}},
  \bibnamefont{and} \bibinfo{author}{\bibfnamefont{M.~S.}
  \bibnamefont{Sarandy}}, \bibinfo{journal}{Physical Review A}
  \textbf{\bibinfo{volume}{87(6)}}, \bibinfo{pages}{064101}
  (\bibinfo{year}{2013}).

\bibitem[{\citenamefont{Dakic et~al.}(2010)\citenamefont{Dakic, Vedral, and
  Brukner}}]{64}
\bibinfo{author}{\bibfnamefont{B.}~\bibnamefont{Dakic}},
  \bibinfo{author}{\bibfnamefont{V.}~\bibnamefont{Vedral}}, \bibnamefont{and}
  \bibinfo{author}{\bibfnamefont{C.}~\bibnamefont{Brukner}},
  \bibinfo{journal}{Physical review letters}
  \textbf{\bibinfo{volume}{105(19)}}, \bibinfo{pages}{190502}
  (\bibinfo{year}{2010}).

\bibitem[{\citenamefont{Nakano et~al.}(2013{\natexlab{a}})\citenamefont{Nakano,
  Piani, and Adesso}}]{65}
\bibinfo{author}{\bibfnamefont{T.}~\bibnamefont{Nakano}},
  \bibinfo{author}{\bibfnamefont{M.}~\bibnamefont{Piani}}, \bibnamefont{and}
  \bibinfo{author}{\bibfnamefont{G.}~\bibnamefont{Adesso}},
  \bibinfo{journal}{Physical Review A} \textbf{\bibinfo{volume}{88(1)}},
  \bibinfo{pages}{012117} (\bibinfo{year}{2013}{\natexlab{a}}).

\bibitem[{\citenamefont{Ma et~al.}(2015)\citenamefont{Ma, Chen, Fanchini, and
  Fei}}]{66}
\bibinfo{author}{\bibfnamefont{Z.}~\bibnamefont{Ma}},
  \bibinfo{author}{\bibfnamefont{Z.}~\bibnamefont{Chen}},
  \bibinfo{author}{\bibfnamefont{F.}~\bibnamefont{Fanchini}}, \bibnamefont{and}
  \bibinfo{author}{\bibfnamefont{S.~M.} \bibnamefont{Fei}},
  \bibinfo{journal}{Scientific reports} \textbf{\bibinfo{volume}{5(1)}},
  \bibinfo{pages}{10262} (\bibinfo{year}{2015}).

\bibitem[{\citenamefont{Obando et~al.}(2015)\citenamefont{Obando, Paula, and
  Sarandy}}]{67}
\bibinfo{author}{\bibfnamefont{P.}~\bibnamefont{Obando}},
  \bibinfo{author}{\bibfnamefont{F.~M.} \bibnamefont{Paula}}, \bibnamefont{and}
  \bibinfo{author}{\bibfnamefont{M.~S.} \bibnamefont{Sarandy}},
  \bibinfo{journal}{Physical Review A} \textbf{\bibinfo{volume}{92(3)}},
  \bibinfo{pages}{032307} (\bibinfo{year}{2015}).

\bibitem[{\citenamefont{Luo}(2008)}]{68}
\bibinfo{author}{\bibfnamefont{S.}~\bibnamefont{Luo}},
  \bibinfo{journal}{Physical Review A} \textbf{\bibinfo{volume}{77(4)}},
  \bibinfo{pages}{042303} (\bibinfo{year}{2008}).

\bibitem[{\citenamefont{Khedif et~al.}(2022{\natexlab{b}})\citenamefont{Khedif,
  Haddadi, Daoud, Dolatkhah, and Pourkarimi}}]{69}
\bibinfo{author}{\bibfnamefont{Y.}~\bibnamefont{Khedif}},
  \bibinfo{author}{\bibfnamefont{S.}~\bibnamefont{Haddadi}},
  \bibinfo{author}{\bibfnamefont{M.}~\bibnamefont{Daoud}},
  \bibinfo{author}{\bibfnamefont{H.}~\bibnamefont{Dolatkhah}},
  \bibnamefont{and} \bibinfo{author}{\bibfnamefont{M.~R.}
  \bibnamefont{Pourkarimi}}, \bibinfo{journal}{Quantum Information Processing}
  \textbf{\bibinfo{volume}{21(7)}}, \bibinfo{pages}{235}
  (\bibinfo{year}{2022}{\natexlab{b}}).

\bibitem[{\citenamefont{Nakano et~al.}(2013{\natexlab{b}})\citenamefont{Nakano,
  Piani, and Adesso}}]{70}
\bibinfo{author}{\bibfnamefont{T.}~\bibnamefont{Nakano}},
  \bibinfo{author}{\bibfnamefont{M.}~\bibnamefont{Piani}}, \bibnamefont{and}
  \bibinfo{author}{\bibfnamefont{G.}~\bibnamefont{Adesso}},
  \bibinfo{journal}{Physical Review A} \textbf{\bibinfo{volume}{88(1)}},
  \bibinfo{pages}{012117} (\bibinfo{year}{2013}{\natexlab{b}}).

\bibitem[{\citenamefont{Ciccarello et~al.}(2014)\citenamefont{Ciccarello,
  Tufarelli, and Giovannetti}}]{71}
\bibinfo{author}{\bibfnamefont{F.}~\bibnamefont{Ciccarello}},
  \bibinfo{author}{\bibfnamefont{T.}~\bibnamefont{Tufarelli}},
  \bibnamefont{and} \bibinfo{author}{\bibnamefont{Giovannetti}},
  \bibinfo{journal}{New Journal of Physics} \textbf{\bibinfo{volume}{16(1)}},
  \bibinfo{pages}{013038} (\bibinfo{year}{2014}).

\end{thebibliography}

\bigskip

\end{document}